\documentclass[useAMS,usenatbib]{mn2e} 
\usepackage{graphicx}
\usepackage{color}
\usepackage{mathrsfs}
\usepackage{txfonts}
\usepackage{epstopdf}

   \title[Chemical enrichment of Bo\"otes~I]{Chemical enrichment in very 
     low-metallicity environments: Bo\"otes~I}
   \author[D.~Romano et al.]{Donatella Romano,$^{1}$\thanks{E-mail: 
       donatella.romano@oabo.inaf.it} Michele Bellazzini,$^{1}$ Else 
     Starkenburg,$^{2}$\thanks{CIfAR Global Scholar.} and Ryan Leaman$^{3,4}$\\
     $^{1}$INAF, Osservatorio Astronomico di Bologna, Via Ranzani 1, I-40127 
     Bologna, Italy\\
     $^{2}$Department of Physics and Astronomy, University of Victoria, PO Box 
     3055 STN CSC, Victoria, BC V8W 3P6, Canada\\
     $^{3}$Instituto de Astrof{\'\i}sica de Canarias, E-38205 La Laguna, 
     Tenerife, Spain\\
     $^{4}$Departamento de Astrof{\'\i}sica, Universidad de La Laguna, E-38205 
     La Laguna, Tenerife, Spain}

\begin{document}

     \date{Accepted 2014 November 13. Received 2014 November 5; 
       in original form 2014 July 29}

     \pagerange{\pageref{firstpage}--\pageref{lastpage}} \pubyear{2014}

     \maketitle

     \label{firstpage}


   \begin{abstract}
     We present different chemical evolution models for the ultrafaint dwarf 
     galaxy Bo\"otes~I. We either assume that the galaxy accretes its mass 
     through smooth infall of gas of primordial chemical composition ({\it 
       classical} models) or adopt mass accretion histories derived from the 
     combination of merger trees with semi-analytical modelling ({\it 
       cosmologically-motivated} models). Furthermore, we consider models with 
     and without taking into account inhomogeneous mixing in the ISM within the 
     galaxy, i.e. {\it homogeneous} versus {\it inhomogeneous} models. The 
     theoretical predictions are then compared to each other and to the body of 
     the available data. From this analysis, we confirm previous findings that 
     Bo\"otes~I has formed stars with very low efficiency but, at variance with 
     previous studies, we do not find a clear-cut indication that supernova 
     explosions have sustained long-lasting galactic-scale outflows in this 
     galaxy. Therefore, we suggest that external mechanisms such as ram 
     pressure stripping and tidal stripping are needed to explain the absence 
     of neutral gas in Bo\"otes~I today.
   \end{abstract}

   \begin{keywords}
     galaxies: abundances -- galaxies: dwarf -- galaxies: evolution -- 
     galaxies: individual (Bo\"otes~I) -- stars: abundances.
   \end{keywords}


   \section{Introduction}
   \label{sec:int}

   Since its recent discovery (Belokurov et al. 2006), the Bo\"otes~I 
   ultrafaint dwarf spheroidal galaxy (UFD) has been the subject of a number of 
   investigations, aiming at providing insight into the formation and evolution 
   of this system. Located at 66\,$\pm$\,2 kpc distance from the Sun and with 
   an absolute visual magnitude of $M_V = -$6.3\,$\pm$\,0.2 mag (McConnachie 
   2012), Bo\"otes~I is one of the brightest UFDs found lurking around the 
   Milky Way, and one of its closest satellites. The color-magnitude diagrams 
   reveal that its stellar population is old and metal-poor, with age spread 
   (if any) limited to a few billion years (de Jong et al. 2008; Hughes, 
   Wallerstein \& Bossi 2008). No H{\sevensize I} is detected in or around 
   Bo\"otes~I to a 3$\sigma$ upper limit of 180~M$_\odot$ within the optical 
   half-light radius; the resulting H{\sevensize I} mass-to-light ratio is, 
   thus, extremely low, less than 0.002~M$_\odot$/L$_\odot$, which makes it one 
   of the most gas-poor galaxies known  (Bailin \& Ford 2007). Whichever 
   mechanisms are responsible for gas removal from Bo\"otes~I, the absence of 
   gas streams or outflowing gas suggests that they have completed long ago. 
   Bo\"otes~I also exhibits extremely irregular density contours (Belokurov et 
   al. 2006), which indicates that it is undergoing tidal disruption.

   Estimates of the dynamical mass enclosed within the half-light radius ($r_h 
   =$ 242\,$\pm$\,21 pc; McConnachie 2012) have suggested that this extremely 
   low-luminosity system is possibly the darkest Milky Way satellite, with 
   $M_{\mathrm{dyn}}(\le r_h) =$ 2.36\,$\times$\,10$^7$~M$_\odot$, when a stellar 
   velocity dispersion of $\sigma_v =$~9.0~km s$^{-1}$ is adopted (Wolf et al. 
   2010). However, large uncertainties affect this determination and the mass 
   may be sensibly smaller, with a lower limit of 
   8.1\,$\times$\,10$^5$~M$_\odot$ reported by McConnachie (2012), based on 
   Koposov et al.'s (2011) lower value for $\sigma_v$. Koposov et al. actually 
   find two kinematically distinct stellar components in Bo\"otes~I: a 
   dominant, `cold' component, that encompasses 70 per cent of the member stars 
   and has a low projected radial velocity dispersion of 2.4$^{+0.9}_{-0.5}$~km 
   s$^{-1}$, and a minority, `hot' component, that encompasses 30 per cent of 
   the member stars and has a projected radial velocity dispersion of about 
   9~km s$^{-1}$. They speculate that this may arise from the velocity 
   anisotropy of the stellar population.

   Despite its small baryonic mass (the stellar mass is only 
   2.9\,$\times$\,10$^4$ M$_\odot$, assuming a stellar mass-to-light ratio of 
   1; McConnachie 2012), Bo\"otes~I shows a large star-to-star variation in Fe 
   abundances (Norris et al. 2008, 2010a; Feltzing et al. 2009; Lai et al. 
   2011; Gilmore et al. 2013), in starkingly sharp contrast with the lack of 
   evidence for [Fe/H] dispersion in Galactic globular clusters of similar mass 
   (see the review by Gratton, Sneden \& Carretta 2004 and Carretta et al. 
   2009). Iron abundances [Fe/H] ranging from $-$3.7 to $-$1.65 in Bo\"otes~I 
   clearly point to self-enrichment from essentially primordial initial 
   chemical composition and provide indirect evidence for the presence of dark 
   matter (Norris et al. 2008) or, alternatively, for a baryonic mass at the 
   epoch of chemical enrichment significantly higher than seen today.
   

   \begin{table*}
     \setlength{\tabcolsep}{5pt}
     \caption{ High-resolution abundances of Bo\"otes~I stars.}
     \begin{tabular}{@{}lrlrlrlrlrlrlrlc@{}}
       \hline
       Star & [Fe/H] & $\sigma$ & [C/Fe] & $\sigma$ & [Na/Fe] & $\sigma$ & 
       [Mg/Fe] & $\sigma$ & [Ca/Fe] & $\sigma$ & [Ti/Fe] & $\sigma$ & 
       [Ba/Fe] & $\sigma$ & Source/Instr.$^a$ \\
       \hline
       Boo-7        & $-$2.33 & 0.05 &          &      &         &      & 0.41 & 0.11 & 0.23    & 0.11 &         &      & $-$0.75 &      & 1/H \\ 
       Boo-9        & $-$2.64 & 0.14 & $<-$0.29 &      &    0.01 & 0.27 & 0.13 & 0.16 & 0.18    & 0.05 &    0.26 & 0.10 & $-$0.89 & 0.27 & 3/S \\ 
       Boo-33       & $-$2.52 & 0.07 &          &      &         &      & 0.45 & 0.17 & 0.40    & 0.14 &         &      & $-$0.40 & 0.18 & 1/H \\ 
                    & $-$2.32 & 0.16 &          &      &    0.05 & 0.23 & 0.26 & 0.22 & 0.14    & 0.06 & $-$0.02 & 0.17 & $-$0.40 & 0.19 & 2/U \\ 
       Boo-41       & $-$1.88 & 0.16 &          &      &         &      & 0.50 & 0.22 & 0.28    & 0.06 &    0.78 & 0.20 & $-$0.39 & 0.20 & 2/U \\ 
       Boo-94       & $-$2.95 & 0.04 &          &      &         &      & 0.49 & 0.05 & 0.22    & 0.05 &         &      &         &      & 1/H \\ 
                    & $-$2.94 & 0.16 &          &      &    0.15 & 0.20 & 0.49 & 0.16 & 0.30    & 0.06 &    0.26 & 0.13 & $-$0.94 & 0.20 & 2/U \\ 
                    & $-$3.18 & 0.14 &  $<$0.25 &      & $-$0.32 & 0.20 & 0.39 & 0.06 & 0.46    & 0.05 &    0.28 & 0.19 & $-$0.80 & 0.12 & 3/S \\ 
       Boo-117      & $-$2.29 & 0.06 &          &      &         &      & 0.35 & 0.15 & 0.29    & 0.15 &         &      & $-$0.46 & 0.10 & 1/H \\ 
                    & $-$2.18 & 0.16 &          &      & $-$0.05 & 0.24 & 0.18 & 0.22 & 0.20    & 0.06 &    0.14 & 0.13 & $-$0.65 & 0.25 & 2/U \\ 
                    & $-$2.15 & 0.18 & $-$0.79  & 0.35 & $-$0.25 & 0.08 & 0.04 & 0.14 & 0.01    & 0.13 &    0.23 & 0.17 & $-$0.43 & 0.10 & 3/S \\ 
       Boo-119      & $-$3.33 & 0.16 &          &      &    0.73 & 0.23 & 1.04 & 0.22 & 0.46    & 0.18 &    0.80 & 0.28 & $-$1.00 & 0.24 & 2/U \\ 
       Boo-121      & $-$2.44 & 0.04 &          &      &         &      & 0.37 & 0.16 & 0.38    & 0.14 &         &      & $-$0.43 &      & 1/H \\ 
                    & $-$2.49 & 0.19 & $<-$0.24 &      & $-$0.25 & 0.10 & 0.20 & 0.20 & 0.24    & 0.08 & 0.09    & 0.19 & $-$0.63 & 0.20 & 3/S \\ 
       Boo-127      & $-$2.03 & 0.06 &          &      &         &      & 0.71 & 0.09 & 0.02    & 0.10 &         &      & $-$0.64 & 0.31 & 1/H \\ 
                    & $-$2.01 & 0.16 &          &      &         &      & 0.17 & 0.18 & 0.16    & 0.05 &    0.20 & 0.14 & $-$0.69 & 0.29 & 2/U \\ 
                    & $-$1.92 & 0.21 & $-$0.77  & 0.36 & $-$0.18 & 0.15 & 0.11 & 0.07 & 0.19    & 0.10 &    0.15 & 0.15 & $-$0.87 & 0.24 & 3/S \\ 
       Boo-130      & $-$2.32 & 0.16 &          &      &    0.03 & 0.21 & 0.21 & 0.22 & 0.19    & 0.08 &    0.17 & 0.16 & $-$0.54 & 0.23 & 2/U \\ 
       Boo-911      & $-$2.26 & 0.05 &          &      &         &      & 0.29 & 0.13 & 0.40    & 0.06 &         &      & $-$0.56 & 0.08 & 1/H \\ 
                    & $-$2.16 & 0.23 & $-$0.77  & 0.37 & $-$0.28 & 0.08 & 0.35 & 0.13 & $-$0.01 & 0.12 & $-$0.09 & 0.11 & $-$0.64 & 0.15 & 3/S \\ 
       Boo-1137$^b$ & $-$3.66 & 0.11 &    0.26  & 0.2  & $-$0.08 & 0.14 & 0.30 & 0.21 & 0.55    & 0.14 &    0.48 & 0.10 & $-$0.55 & 0.17  & 2/U \\ 
       \hline
       \end{tabular}
       \label{tab:spec}
       \begin{flushleft}
       \emph{Notes.} Identification system by Norris et al. (2008). Solar 
       reference values from Asplund et al. (2009). \\
       $^a$ 1: Feltzing et al. 2009; 2: Gilmore et al. 2013; 3: Ishigaki et al. 
       2014; H: High Resolution Echelle Spectrometer (HIRES) on Keck I; U: 
       Ultraviolet and Visual Echelle Spectrograph (UVES) on VLT; S: High 
       Dispersion Spectrograph (HDS) on Subaru. \\
       $^b$ [Fe/H], [Mg/Fe], [Ca/Fe] and [Ti/Fe] from Table~8 of Gilmore et al. 
       (2013); [C/Fe], [Na/Fe] and [Ba/Fe] from Table~2 of Norris et al. 
       (2010b), placed on the Asplund et al. (2009) scale.
       \end{flushleft}
   \end{table*}

   
   Recently, Vincenzo et al. (2014) modelled the chemical evolution of 
   Bo\"otes~I and concluded that this galaxy experienced very low star 
   formation activity and efficient galactic winds, that got rid of all its 
   gas. Their findings are in agreement with previous claims by Salvadori \& 
   Ferrara (2009) that UFDs have formed stars very ineffectively, turning less 
   than 3 per cent of their baryons into stars.

   All its characteristics make Bo\"otes~I an outstanding local benchmark of 
   the elusive earliest stages of galaxy formation. The purpose of this paper 
   is to interpret the chemical features of stars in Bo\"otes~I in terms of 
   enrichment time-scales and early evolutionary conditions. In Section~2, we 
   review the available spectroscopic data. In Section~3, we present our 
   chemical evolution models. In Section~4 our results are discussed, in 
   comparison to the observations. Finally, in Section~5 some conclusions are 
   drawn.

   \section{Observational data}
   \label{sec:data}

   \subsection{Low- and high-resolution spectroscopy}

   Nowadays, it is widely recognized that the relative distribution of the 
   chemical elements inside galaxies and the spread in abundance ratios are 
   major diagnostics of structure formation and evolution. Reliable abundance 
   determinations of a large number of chemical species are best obtained by 
   means of high spectral resolution analyses. However, high-resolution 
   spectroscopic measurements of individual members of the faintest Milky Way 
   companions are extremely challenging. At low resolution, the exposure times 
   significantly shorten and larger samples of stars can be probed. However, 
   from low-resolution spectra typically only iron, $\alpha$ elements and 
   carbon abundances can be obtained. Medium-resolution spectroscopy 
   ($R \sim$~6000) offers a reasonable trade-off, especially if coupled to 
   multi-object capability, by enabling valuable data to be collected for 
   thousands of stars in Milky Way satellites (e.g. Battaglia et al. 2006; 
   Kirby et al. 2010; Vargas et al. 2013). In the following, we review the 
   spectroscopic data available for the Bo\"otes~I UFD.

   Lai et al. (2011) have presented a chemical abundance analysis of 25 members 
   of Bo\"otes~I. Target stars were selected from the radial velocity survey of 
   Martin et al. (2007) and observed with the Low Resolution Imaging 
   Spectrometer (LRIS; Oke et al. 1995) mounted on the Keck~I telescope. The 
   resolving power is $R \sim$~1800 at 5100 \AA. The data allow measurements of 
   [Fe/H], [C/Fe] and [$\alpha$/Fe] for each star. A significant spread in 
   metallicity (2.1 dex in [Fe/H]) and a low mean iron abundance 
   ($\langle$[Fe/H]$\rangle$=$-$2.59 dex) are found, matching previous 
   estimates (see next paragraph). The authors caution that a systematic offset 
   on the order $-$0.13~dex from literature values is likely present in their 
   [$\alpha$/Fe] ratios.

   High-resolution spectroscopic studies have been carried out for sparse 
   samples of Bo\"otes~I stars, using different instrumentation (see 
   Table~\ref{tab:spec}). Feltzing et al. (2009) have found a mean metallicity 
   of $-$2.3 dex for seven red giant stars with individual [Fe/H] values 
   ranging from $-$2.9 to $-$1.9 dex. One star in their sample, Boo-127, shows 
   an unusually high [Mg/Ca] ratio, similarly to Dra-119 in Draco (Fulbright et 
   al. 2004) and Her-2 and Her-3 in Hercules (Koch et al. 2008). The high value 
   of [Mg/Fe] for Boo-127, however, is not confirmed by Gilmore et al. (2013), 
   who, on the other hand, find excellent abundance agreement for Ca, Ba and Fe 
   for the four stars they have in common with Feltzing et al. (2009). One star 
   in Gilmore et al.'s (2013) sample, Boo-41, displays an anomalously high Ti 
   abundance. Another object, Boo-119, is a carbon-enhanced metal poor star 
   ([Fe/H]~=$-$3.33 dex) with no neutron-capture element enhancement (CEMP-no 
   star) and [Mg/Fe]~=~1.04 dex. If both these stars are excluded, a weak 
   signature of declining [$\alpha$/Fe] with increasing [Fe/H] is found, in 
   contrast to what is known for field halo stars of the same metallicities 
   (see also Vargas et al. 2013). The chemical evolution of Bo\"otes~I is 
   suggested to have proceeded in a homogeneous manner, at variance with 
   Feltzing et al. (2009), who support the presence of inhomogeneities. In 
   order to shed light on these discrepancies, Ishigaki et al. (2014) have 
   recently performed an independent analysis of six red giant stars in 
   Bo\"otes~I. They have five stars and three stars, respectively, in common 
   with Feltzing et al. (2009) and Gilmore et al. (2013). Their analysis does 
   not support a highly inhomogeneous chemical evolution for Bo\"otes~I and 
   suggests a low value for [Mg/Fe] in Boo-127, fully consistent with Gilmore 
   et al. (2013) and in disagreement with Feltzing et al. (2009). Finally, 
   carbon abundances from medium-resolution spectra for sixteen Bo\"otes~I red 
   giants are presented in Norris et al. (2010a). The spread in carbon 
   abundances is large, $\Delta$[C/H]~=~1.5 dex.

   For the purpose of the present work, we adopt the high-resolution data 
   published in Feltzing et al. (2009), Norris et al. (2010b), Gilmore et al. 
   (2013) and Ishigaki et al. (2014) (see Table~\ref{tab:spec}, where the 
   ratios have been adjusted using Asplund et al. 2009 solar reference values 
   if needed). Moreover, we adopt carbon abundances by Norris et al. (2010a) 
   and Lai et al. (2011) for sixteen stars and one star (Boo-119), 
   respectively. These abundances have been inferred from medium- and 
   low-resolution spectra and are not listed in Table~\ref{tab:spec}. To ensure 
   a good statistic, we also use as a constraint the Bo\"otes~I `combined' 
   metallicity distribution function (MDF) obtained by Lai et al. (2011) by 
   expanding their sample to include non-overlapping stars from Norris et al. 
   (2010a) and Feltzing et al. (2009); the resulting empirical distribution 
   totals 41 stars.

   \section{Chemical evolution models}
   \label{sec:model}


   \begin{table*}
     \caption{ Parameters and final properties of the classical homogeneous 
       models.}
     \begin{tabular}{@{}lccccccccccccl@{}}
       \hline
       Model & ${\mathscr M}_{{\mathrm{b}}}$ & $\tau$ & $\nu$ & 
       $\Delta t_{\mathrm{SF}}$ & $r_h/r_{\mathrm{DM}}$ & 
       ${\mathscr M}_{\mathrm{DM}}$ & $\varepsilon$ & $w_{\mathrm{heavy}}$ & 
       ${\mathscr M}_{\mathrm{stars}}$ & $\langle$[Fe/H]$\rangle_{\mathrm{stars}}$ 
       & $\Delta t_{\mathrm{out}}$ & ${\mathscr M}_{\mathrm{gas}}$ & Flag \\
       & (M$_\odot$) & (Myr) & (Gyr$^{-1}$) & (Gyr) & & (M$_\odot$) & & & 
       (10$^4$ M$_\odot$) & (dex) & (Myr) & (10$^6$ M$_\odot$) & \\
       \hline
       Boo\,{\sevensize 1}     & 2\,$\times$\,10$^6$              & 50 & 0.02 & 1    & 0.1 & 2\,$\times$\,10$^6$ & 0.01  & 12 & 1.5 & $-$2.24 & 285 & 1.87 & \\
       Boo\,{\sevensize 2}     & 2\,$\times$\,10$^6$              & 50 & 0.02 & 1    & 0.1 & 2\,$\times$\,10$^6$ & 0.1 & 12 & 1.5 & $-$2.25 & 45 & 1.74 & D \\
       Boo\,{\sevensize 3}     & 2\,$\times$\,10$^6$              & 50 & 0.02 & 1    & --- & --- & 0.01  & 12 & 1.5 & $-$2.24 & 275 & 1.87 & \\
       Boo\,{\sevensize 4}     & 2\,$\times$\,10$^6$              & 50 & 0.04 & 1    & 0.1 & 2\,$\times$\,10$^6$ & 0.01  & 12 & 3.0 & $-$1.92 & 155  & 1.84 & \\
       Boo\,{\sevensize 5}     & 1.1\,$\times$\,10$^7$ & 50 & 0.013 & 1    & 0.1 & 1.1\,$\times$\,10$^8$ & 0.1  & 12 & 5.5 & $-$2.44 & 310 & 10.4 & \\
       Boo\,{\sevensize 6}     & 1.1\,$\times$\,10$^7$ & 50 & 0.026 & 0.5  & 0.1 & 1.1\,$\times$\,10$^8$ & 0.1  & 12 & 5.5 & $-$2.52 & 165 & 10.4 & \\
       Boo\,{\sevensize 7}     & 1.1\,$\times$\,10$^7$ & 50 & 0.053 & 0.25 & 0.1 & 1.1\,$\times$\,10$^8$ & 0.01  & 12 & 5.5 & $-$2.60 & --- & 10.4 & \\
       Boo\,{\sevensize 8}$^a$ & 1.1\,$\times$\,10$^7$ & 50 & 0.053 & 0.25 & 0.1 & 1.1\,$\times$\,10$^8$ & 0.01  & 12 & 5.5 & $-$2.60 & --- & 10.4 & \\
                               & 10$^5$              & 50 & 4.0  & 0.02 & 0.1 & 1.1\,$\times$\,10$^8$ & 0.01  & 12 & 0.25 & $-$2.46 & 15  & 0.09 & D \\
       Boo\,{\sevensize 9}     & 1.1\,$\times$\,10$^7$ & 50 & 0.013 & 1    & 0.1 & 6.5\,$\times$\,10$^7$ & 0.1  & 12 & 5.5 & $-$2.44 & 280 & 10.3 & \\
       \hline
       \end{tabular}
       \label{tab:mod}
       \begin{flushleft}
       \emph{Notes.} Listed in columns 1 to 13 are: the model name; the total 
       baryonic (gaseous) mass accreted by the system; the infall time-scale; 
       the star formation efficiency; the duration of the star formation 
       episode; the ratio of the effective-to-dark matter core radius; the 
       total dynamical mass of the system; the thermalization efficiency from 
       SNe of all types and stellar winds; the efficiency of metal removal from 
       the star forming regions; the final stellar mass of the system; the 
       average stellar metallicity; the time interval between the beginning of 
       star formation and the onset of the first episode of gas removal from 
       the star forming region (that does not necessarily lead to gas removal 
       from the system); the final gaseous mass of the system. A capital `D' in 
       the last column denotes whether the energy output from SNe exceeds the 
       binding energy of the galaxy (which results in disrupting the object).\\
       $^a$ This model is characterized by two bursts of star formation. The 
       model parameters (columns 2--9) and final properties (columns 10--13) 
       corresponding to the two distinct star formation episodes are listed in 
       distinct rows.
       \end{flushleft}
   \end{table*}


   We compute the evolution of the abundances of several chemical elements (H, 
   D, He, Li, C, N, O, Na, Mg, Al, Si, S, Ca, Sc, Ti, Cr, Mn, Co, Ni, Fe, Cu, 
   Zn) in the interstellar medium (ISM) of Bo\"otes~I. We use detailed 
   numerical models, that solve the classical set of equations of chemical 
   evolution (see e.g. Tinsley 1980; Pagel 1997; Matteucci 2001, 2012). 

   \subsection{Classical models}

   Our {\it classical} models rest on the following assumptions:
   \begin{enumerate}
     \item inflow of gas of primordial chemical composition provides the raw 
       material for star formation;
     \item galactic outflows remove gas from the system;
     \item the stellar initial mass function (IMF) is constant in space and 
       time;
     \item the finite stellar lifetimes are taken into account (no 
       {\it instantaneous recycling approximation,} IRA, is adopted).
   \end{enumerate}

   \subsubsection{Gas accretion and star formation}

   For the sake of simplicity, the rate of gas infall is parametrized as
   \begin{equation}
     \frac{{\mathrm d}{\mathscr M}_{\mathrm{b}}}{{\mathrm d}t} \propto
     {\mathrm e}^{-t/\tau},
     \label{eq:in}
   \end{equation}
   where ${\mathscr M}_{\mathrm{b}}$, the total amount of matter ever accreted, 
   and $\tau$, the e-folding time-scale, are free parameters of the models. A 
   smooth accretion of gas is clearly a rough approximation of the true 
   assembly history of galaxies. According to modern theories of galaxy 
   formation, in fact, the mass assembly must proceed through discrete episodes 
   and mergers play a key role (see e.g. Conselice 2012, for a recent review). 
   However, if most of the mass is gathered early on in gaseous subclumps and 
   the stars form mostly {\it in situ}, the chemical properties predicted for 
   the bulk of the stellar population are expected to be quite robust against 
   the simplified mass assembly history implied by Equation~(\ref{eq:in}). 

   The gas is turned into stars following a Kennicutt-Schmidt law (Schmidt 
   1959; Kennicutt 1998):
   \begin{equation}
     \psi(t) = \nu {\mathscr M}_{\mathrm{gas}}(t),
   \end{equation}
   where $\nu$, the efficiency of the process, is a free parameter of the 
   models. To be precise, the original Kennicutt (1998) law refers to surface 
   densities: $\dot{\Sigma}_{\star} \propto \Sigma_{\mathrm{gas}}^{1.4 \pm 0.15}$. For 
   star-forming regions with roughly constant scale heights, the surface 
   densities can be turned into volume densities: $\dot{\rho_{\star}} \propto 
   \rho_{\mathrm{gas}}^{1.5}$. By assuming that the Kennicutt-Schmidt law indicates 
   that the star formation rate is controlled by the self-gravity of the gas, 
   one can write: $\dot{\rho_{\star}} = \varepsilon 
   \rho_{\mathrm{gas}}/t_{\mathrm{ff}}$, where $t_{\mathrm{ff}} \propto 
   1/\sqrt{\rho_{\mathrm{gas}}}$ and $\varepsilon$ is a free parameter. We adopt 
   this formulation of the star formation rate, with $\nu = 
   \varepsilon/t_{\mathrm{ff}}$. We do not consider a gas density threshold for 
   star formation. As for the stellar IMF, we adopt a Kroupa (2001) IMF in the 
   mass range 0.1--100 M$_\odot$, unless otherwise stated.

   \subsubsection{Mechanical feedback}

   Thermal feedback from stars is included by assuming that type II and type Ia 
   supernovae (SNe~II and SNe~Ia, respectively) deposit $E_{\mathrm{SN}} =$ 
   10$^{51}$ erg of energy each\footnote{ The consequences of the occurrence of 
     hypernovae with more energetic outputs are not explored.} into the ISM. 
   The energy injected by a typical massive star via stellar winds during its 
   lifetime ($E_{\mathrm{wind}} =$ 10$^{49}$ erg; see Gibson 1994, his figure~1) is 
   added though it has a negligible effect on the global energy budget. In 
   order to account for radiative energy losses, we assume constant values of 
   the thermalization efficiencies, $\varepsilon_{\mathrm{SNII}} = 
   \varepsilon_{\mathrm{SNIa}} = \varepsilon_{\mathrm{wind}} = \varepsilon =$ 
   0.01--0.1 (see Table~\ref{tab:mod}). It is worth stressing that there is not 
   general consensus about these values in the literature (see e.g. Recchi 
   2014, for a recent review). Furthermore, the heating efficiency is likely to 
   be a time-dependent quantity (Melioli \& de Gouveia Dal Pino 2004).

   At each time step, we compute the gas thermal and binding energies following 
   Bradamante, Matteucci \& D'Ercole (1998). When the thermal energy of the gas 
   exceeds its binding energy, an outflow develops and the thermal energy of 
   the gas is reset to zero. The rate of gas loss through the outflow is 
   assumed to be proportional to the star formation rate
   \begin{equation}
     \frac{{\mathrm d}{\mathscr M}^{\mathrm{out}}}{{\mathrm d}t} = w \,
     \psi(t),
     \label{eq:out}
   \end{equation}
   where $w$ is a further free parameter of the models describing the 
   efficiency of the galactic wind. Following both theoretical and empirical 
   considerations (e.g. Vader 1986, 1987; Recchi, Matteucci \& D'Ercole 2001; 
   Martin, Kobulnicky \& Heckman 2002) we assume a higher ejection efficiency 
   for the heavy elements freshly synthesised in SN explosions than for the 
   neutral ISM ({\it selective winds}; Marconi, Matteucci \& Tosi 1994); in 
   particular, we assume $w_{\mathrm{heavy}} =$ 2~$w_{\mathrm{H, He}}$~= 12. 
   
   The final fate of the swept-up gas and supernova ejecta is matter of debate. 
   Silich \& Tenorio-Tagle (1998) find that the gas remains bound in the hot 
   gaseous halos surrounding galaxies as massive as 10$^9$--10$^{10}$ M$_\odot$. 
   A similar conclusion is reached by Marcolini et al. (2006) for lower mass 
   galaxies (they analyse the specific case of  Draco, a classical dwarf 
   spheroidal galaxy of the Local Group). In this picture, gas removal 
   ultimately results from ram pressure stripping (Mori \& Burkert 2000) and/or 
   tidal interaction with the Milky Way (Mayer et al. 2006). We note that our 
   more massive models have low circular velocity, $V_\mathrm{c} =$ 14~km 
   s$^{-1}$,\footnote{ Following Desai et al. (2004), this corresponds to a 
     stellar central velocity dispersion $\sigma_v =$ 9~km s$^{-1}$, that agrees 
     well with the observed values (2.4--9.0~km s$^{-1}$, depending on the 
     considered component, see Koposov et al. 2011).} and virial temperature 
   below 10$^4$~K. In such conditions, gas cools inefficiently (Sutherland \& 
   Dopita 1993). The cooling time is expected to be (slightly) lower than the 
   (out)flow time and the material entrained in the outflow can possibly leave  
   the galaxy (see Wang 1995). Clearly, the arguments above suffer of important 
   oversimplifications --for instance, we do not take into account the role of 
   gas geometry on the development of galactic winds (the interested reader is 
   referred to Recchi \& Hensler 2013 for a recent reappraisal of this 
   problem). For the purpose of the present work, it suffices to say that in 
   the framework of our classic chemical evolution models the gas entrained in 
   the outflow is assumed to never re-enter the star-forming regions. Sawala et 
   al. (2010, 2014) provide a more realistic treatment of the gas in dwarf 
   galaxies with a range of masses that partly overlaps the one analysed here.

   \subsubsection{Chemical feedback}
   \label{sec:chfdb}

   One of the most important ingrendients of chemical evolution models are, of 
   course, the stellar yields. In this work, we test two different grids of 
   yields for both low- and intermediate-mass stars (1~$\le m$/M$_\odot \le$~8) 
   and massive stars ($m >$ 8~M$_\odot$):
   \begin{enumerate}
     \item the metallicity-dependent yields from van den Hoek \& Groenewegen 
       (1997) for single low- and intermediate-mass stars and those from 
       Woosley \& Weaver (1995, their case B) for core-collapse SNe (our 
       standard choice);
     \item the metallicity-dependent yields from Karakas (2010) for single 
       low- and intermediate-mass stars and those from Kobayashi et al. (2006) 
       for core-collapse SNe.
   \end{enumerate}
   Both grids of yields allow to reproduce reasonably well most of the 
   abundance data for solar neighbourhood stars (see Romano et al. 2010). In 
   Section~\ref{sec:res} we show the results of the models with our standard 
   choice of stellar yields. The impact of adopting different nucleosynthesis 
   prescriptions is discussed in Appendix~A. The yields for SNeIa are always 
   taken from Iwamoto et al. (1999, their model W7). The rate of SNeIa is 
   computed following the recipes outlined in Matteucci \& Greggio (1986).

   The adopted values of the parameters for our classical models are listed in 
   columns 2 to 9 of Table~\ref{tab:mod}.

   \subsection{Cosmologically-motivated models}
   \label{sec:cosmo}

   In the classical chemical evolution modelling scheme described in the 
   previous section, Bo\"otes~I is assumed to form via accretion of gaseous 
   matter of primordial chemical composition, following (rather arbitrarily) a 
   time-decaying infall rate. Possible interactions with the surroundings are 
   not taken into account. However, nowadays the formation sites of even the 
   smallest Milky Way satellites are directly resolved in large cosmological 
   $N$-body dark matter simulations. This enables us to study the formation and 
   evolution of Bo\"otes~I in a $\Lambda$ cold dark matter ($\Lambda$CDM) 
   universe.
    
   We rest on recent work by Starkenburg et al. (2013), who combined six 
   high-resolution Aquarius dark matter simulations (see Springel et al. 2008) 
   with a semi-analytic model of galaxy formation (Li, De Lucia \& Helmi 2010, 
   that is in turn build upon Kauffmann et al. 1999; Springel et al. 2001; De 
   Lucia, Kauffmann \& White 2004; Croton et al. 2006; De Lucia \& Blaizot 
   2007; De Lucia \& Helmi 2008) to investigate the properties of the 
   satellites of Milky Way-like galaxies in a fully cosmological setting (see 
   also, e.g., Cooper et al. 2010; Font et al. 2011; De Lucia et al. 2014, for 
   work on satellites based on the Aquarius Project). In their simulations, new 
   prescriptions are included to follow stellar stripping and tidal disruption 
   of satellites. While the main focus is on the star formation histories of 
   dwarf galaxies in and around the Milky Way, the treatment of chemical 
   enrichment has some limitations due to the adoption of IRA.

   For the purpose of the present work, we identify Bo\"otes~I analogues in the 
   mock galaxy catalogue of Starkenburg et al. (2013), based on rather loose 
   selection criteria: (i) $V$-band magnitude between $-$6.0 and $-$6.5 mag; 
   (ii) mostly old stellar populations (more than 80 per cent of the stars are 
   older than 10~Gyr); (iii) the galaxy is a satellite of the Milky-Way-like 
   halo and no further out than 150~kpc today. We adopt the mass assembly and 
   star formation histories of the Bo\"otes~I candidates and apply the 
   post-processing technique described in Romano \& Starkenburg (2013) in order 
   to obtain the detailed chemical composition of synthetic Bo\"otes~I stars in 
   a fully cosmological framework. We find that for such a small galaxy not all 
   cosmological models are suitable for this post-processing scheme. This is at 
   least partly due to the assumption of IRA in the semi-analytical model, 
   which leads to some inconsistencies with the post-processing code, where 
   this approximation is relaxed. Out of the seventeen selected candidates, we 
   use in this work the eleven for which the post-processing technique can be 
   ran self-consistently. The eleven galaxies that are used in the remainder of 
   this work do represent the full range of star formation histories found in 
   the Bo\"otes~I candidates. For all of these, the totality of the stars are 
   formed \emph{in situ}, i.e. there are not stars brought in the system 
   through mergers. Mergers only provide gas to the system.

   \subsection{Inhomogeneous mixing}
   \label{sec:inho}

   Most chemical evolution studies assume that the ejecta from dying stars are 
   instantly cooled and mixed back into the ambient medium ({\it instantaneous 
     mixing approximation,} IMA). Recchi et al. (2001) have shown that, indeed, 
   most of the metals cool off in a few Myr if a low heating efficiency of a 
   few per cent is adopted for SNe~II. A rapid cooling of the metals would 
   favor an efficient mixing with the ISM (but see Roy \& Kunth 1995; 
   Tenorio-Tagle 1996; Rieschick \& Hensler 2003). In small systems, however, 
   the stochastic sampling of the IMF introduced by the low star formation 
   rates may lead to an internal dispersion in abundance ratios (Carigi \& 
   Hernandez 2008; Cescutti 2008).

   In this study, chemical inhomogeneities are implemented by taking into 
   account the empirical evidence that a strong correlation exists between the 
   mean linear metallicity $\bar{Z}$ and the intrinsic metallicity spread 
   $\sigma(Z)^2$ of Local Group dwarf galaxies and Galactic star clusters 
   (Leaman 2012). At each time step, forming stars do not have all the same 
   mean ISM metallicity, $\bar{Z}$. Rather, they follow a Gaussian distribution 
   in $Z$ values with
   \begin{equation}
     \log \sigma(Z)^2 = a + b \, \log(\bar{Z}),
   \end{equation}
   where $a = -$0.6888970 and $b =$~1.88930 are derived from the slope of the 
   relation for dwarf galaxies in figure~2 of Leaman (2012). The obvious 
   boundary condition applies that stars do not form with negative 
   metallicities. The detailed chemical composition is then obtained by scaling 
   to the mean chemical mixture.

   With this approach, we do not have to introduce further free parameters in 
   order to deal with poorly known physical processes. Detailed 
   three-dimensional hydrodynamical simulations are underway, that will allow 
   us to better quantify the extent of chemical inhomogeneities in Bo\"otes~I 
   and other UFDs in future papers.

   \section{Results}
   \label{sec:res}


   \begin{figure}
     \begin{center}
     \includegraphics[width=\columnwidth]{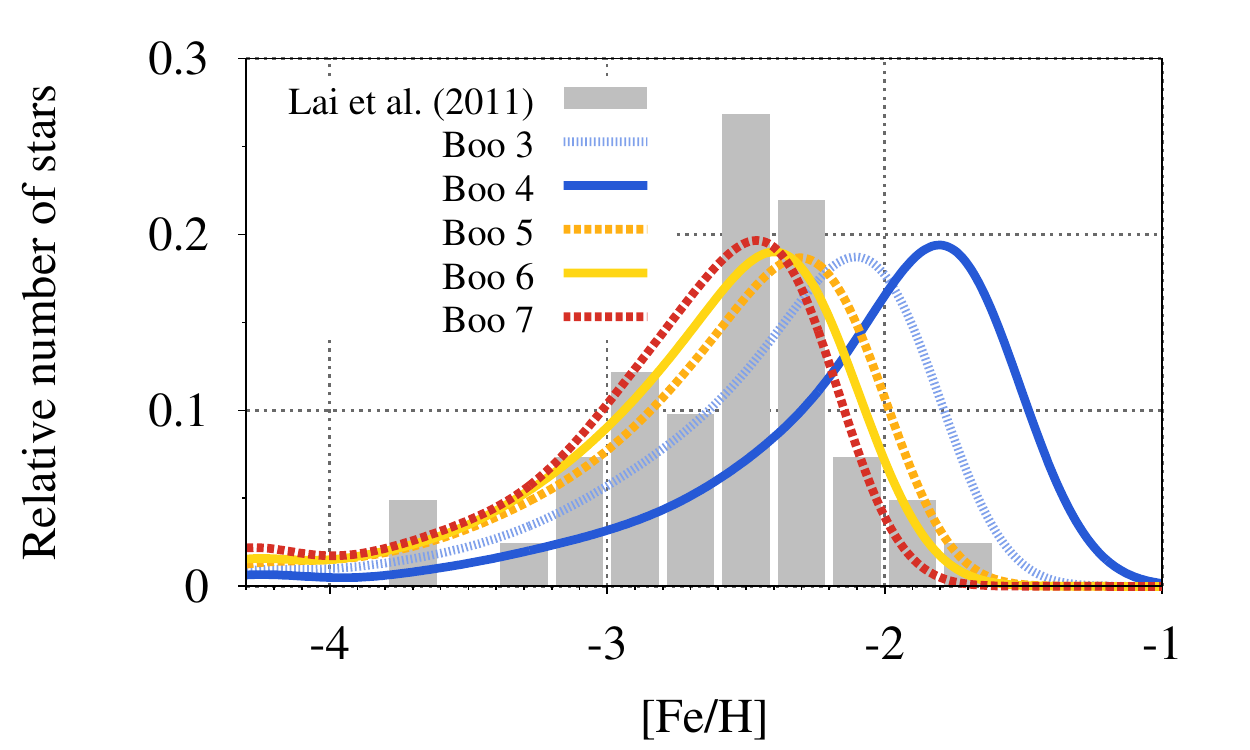}
     \caption{ Theoretical (solid curves) and observed (histogram) MDFs of 
       Bo\"otes~I. The observed MDF is the `combined' MDF presented in Lai et 
       al. (2011). The theoretical MDFs have been convolved with a Gaussian 
       smoothing kernel of $\sigma$= 0.20 dex, to take into account the random 
       individual errors in [Fe/H] determinations.}
     \label{fig:mdfhom}
     \end{center}
   \end{figure}



   \begin{figure}
     \begin{center}
     \includegraphics[width=\columnwidth]{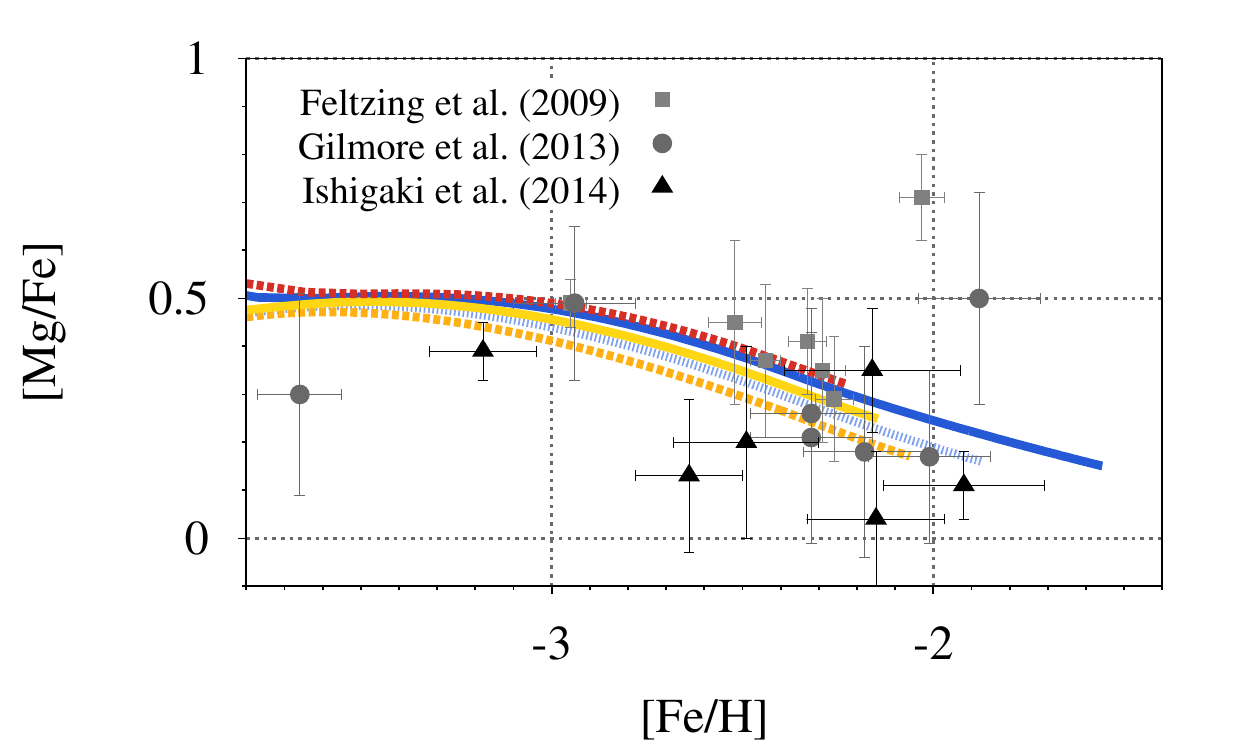}
     \caption{ Theoretical [Mg/Fe] versus [Fe/H] for selected homogeneous 
       chemical evolution models of Bo\"otes~I (coloured lines, colour coding 
       is the same as in Fig.~\ref{fig:mdfhom}). Symbols refer to available 
       high-resolution data (see Table~\ref{tab:spec}).}
     \label{fig:mgfehom}
     \end{center}
   \end{figure}


   In this section we discuss the model results in comparison with the 
   available data. Furthermore, we provide some testable predictions, to be 
   confirmed or disproved by future observations.
 
   \subsection{Homogeneous models}

   Within the homogeneous chemical evolution paradigm, we can address only 
   {\it average} galaxy properties. Nevertheless, homogeneous models turn out 
   to be very useful: because of the small computational demands, they allow a 
   quick screening of the full parameter space, thus paving the way for more 
   sophisticated, computationally expensive modelling. For the classical dwarf 
   spheroidals --and also some UFDs-- of the Local Group, it has been shown 
   that the position of the peak of the stellar MDF, as well as the shape of 
   its wings and the behaviour of the mean [$\alpha$/Fe] ratio as a function of 
   [Fe/H], sensibly constrain the star formation efficiency, gas accretion 
   rate, IMF and occurrence of galactic winds in each galaxy (e.g. Lanfranchi 
   \& Matteucci 2004; Kirby et al. 2011; Vincenzo et al. 2014). The same 
   observational constraints are used in the present study to discriminate 
   among different scenarios for the formation and evolution of Bo\"otes~I.

   \subsubsection{Classical models}


   \begin{figure}
     \begin{center}
     \includegraphics[width=\columnwidth]{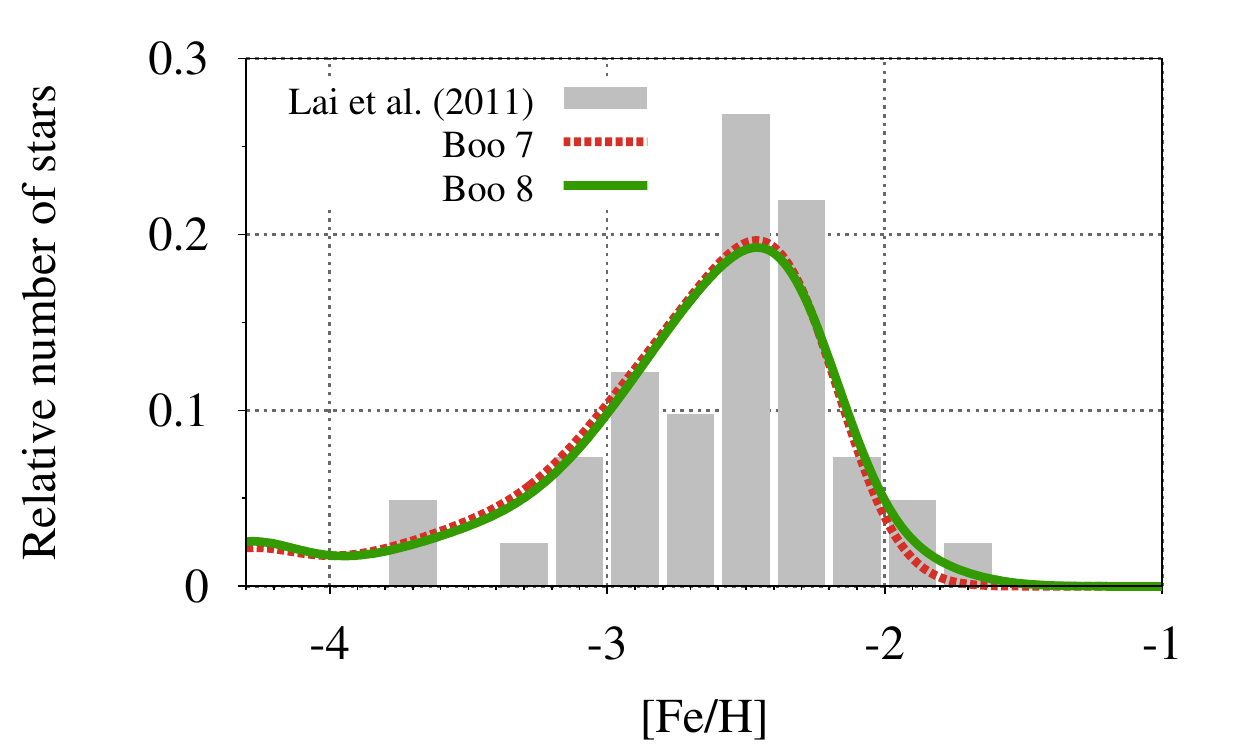}
     \caption{ Same as Fig.~\ref{fig:mdfhom}, for models ~Boo\,{\sevensize 7} 
       (red dashed line) and Boo\,{\sevensize 8} (green solid line). 
       Model~Boo\,{\sevensize 8} is the same as model~Boo\,{\sevensize 7}, 
       except for a secondary star formation burst, extremely short but highly 
       efficient, that is assumed to form a minority population of 
       high-[$\alpha$/Fe], high metallicity stars.}
     \label{fig:mdfhom2}
     \end{center}
   \end{figure}



   \begin{figure}
     \begin{center}
     \includegraphics[width=\columnwidth]{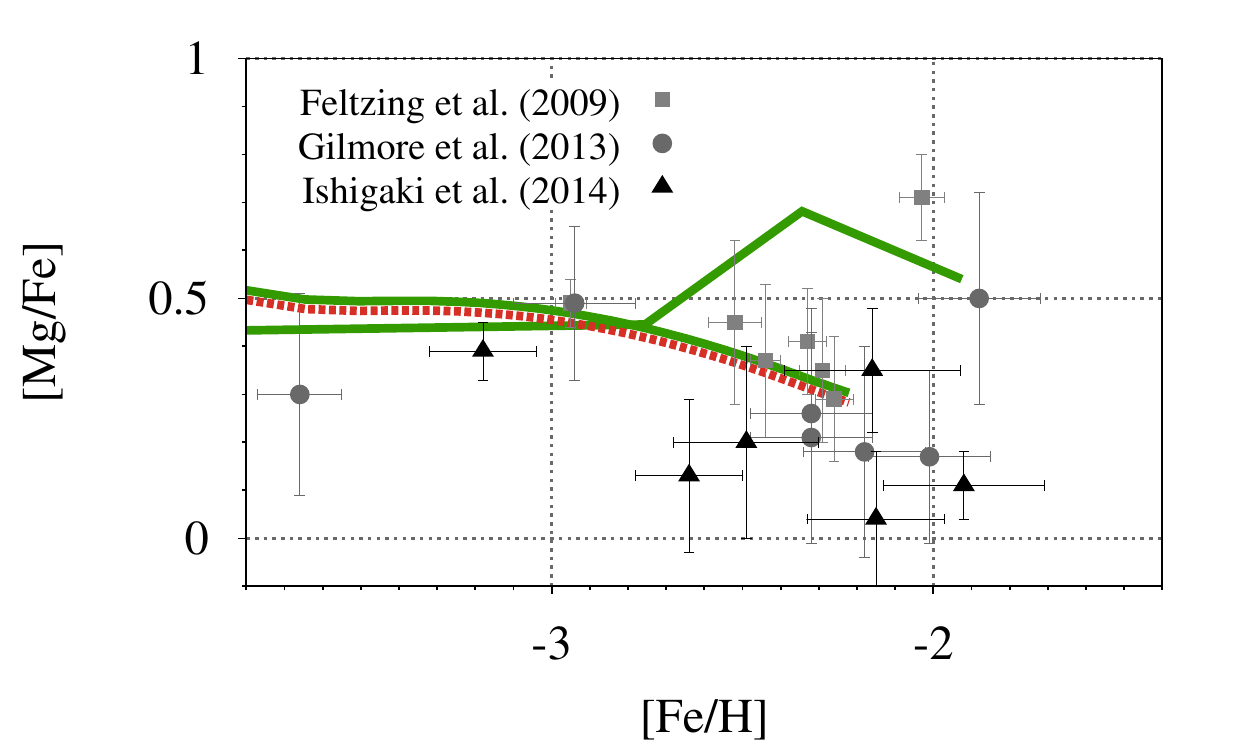}
     \caption{ Same as Fig.~\ref{fig:mgfehom} for models ~Boo\,{\sevensize 7} 
       (red dashed line) and Boo\,{\sevensize 8} (green solid lines; each curve 
       refers to a distinct star formation episode). The predictions of 
       model~Boo\,{\sevensize 7} have been shifted downward by 0.02~dex to make 
       them clearly visible.}
     \label{fig:mgfehom2}
     \end{center}
   \end{figure}


   The adopted values of the input parameters for a subset of selected 
   classical models are listed in Table~\ref{tab:mod} (columns 2 to 9), 
   together with some of the model results (columns 10 to 13). We changed one 
   by one the most important parameters of the simulation: 

   \begin{enumerate}
     \item As reviewed in the Introduction, current estimates of the dynamical 
       mass of Bo\"otes~I enclosed within a half-light radius set lower limits 
       to the total mass of the system that differ by more than one order of 
       magnitude. We consider different amounts of dark matter in the models, 
       ranging from null (model~Boo\,{\sevensize 3}) to $\sim$10$^8$~M$_\odot$ 
       (models~Boo\,{\sevensize 5}--Boo\,{\sevensize 8}). In general, we find 
       that the lower the dark matter content, the higher the probability that 
       SN explosions\footnote{The number of SN explosions in the system is 
         fixed by the observed stellar mass through a fiducial IMF.} destroy 
       the system (this happens when SN explosions unbind all the gas), unless 
       extremely low thermalization efficiencies of about 1 per cent are 
       invoked (cf. models~Boo\,{\sevensize 1} and ~Boo\,{\sevensize 2}). 
       However, our model~Boo\,{\sevensize 3} --with zero dark matter content-- 
       does not undergo disruption. This is due to the fact that, owing to its 
       extremely shallow potential well, the gas heated by SN explosions 
       escapes the system very early on ($\Delta t_{\mathrm{out}}$~= 275 Myr; see 
       Table~\ref{tab:mod}, column~12), carrying away a considerable fraction 
       of the thermal energy released by SNe before it can accumulate and 
       destroy the system. While our treatment of SN feedback is clearly 
       oversimplified and no firm conclusions can be drawn basing on the 
       results of a single model, our findings provide some food for thought 
       regarding the capability of small stellar systems without dark matter to 
       survive multiple SN explosions.
     \item At the beginning of the computation, the structure is assigned a 
       baryon fraction, $f_b = 
       {\mathscr M}_{\mathrm{b}}/{\mathscr M}_{\mathrm{DM}}$, varying from the 
       cosmic fraction ($f_b =$ 0.17\,$\pm$\,0.01; Komatsu et al. 2009, 
       model~Boo\,{\sevensize 9}) to $f_b =$~1 (models~Boo\,{\sevensize 1}, 
       Boo\,{\sevensize 2}, Boo\,{\sevensize 4}). This provides the mass to be 
       accreted by infall (see Equation~\ref{eq:in}). The final baryon fraction 
       is $F_b = {\mathscr M}_{\mathrm{stars}}/{\mathscr M}_{\mathrm{DM}}$, with 
       ${\mathscr M}_{\mathrm{stars}}$ and ${\mathscr M}_{\mathrm{DM}}$ to be read 
       from Table~\ref{tab:mod}, columns~10 and 7, respectively. Notice that, 
       in computing the final baryon fraction, we {\it assume} that the galaxy 
       loses all of its gas by some external mechanism, such as tidal stripping 
       and/or ram pressure stripping. We need to resort to an external 
       mechanism (not implemented in the model) since feedback from SNe is not 
       effective in removing all the neutral gas from the galaxy in our 
       simulations, while the observations indicate that there is likely no 
       H{\sevensize I} in Bo\"otes~I today.
     \item Color-magnitude diagrams reveal that the stellar population of 
       Bo\"otes~I is old and has a small age spread. Resting on this piece of 
       evidence, we test star formation histories consisting of one ancient 
       burst, lasting 0.25, 0.5 or 1 Gyr. The star formation efficiency is 
       fixed such as to predict a stellar mass of $\sim$1.5--6\,$\times$10$^4$ 
       M$_\odot$ at the present time: the longer the duration of star formation, 
       the lower must be the star formation efficiency.
     \item Because of the very shallow potential wells considered in this 
       study, only small heating efficiencies (of a few percent at most) can be 
       tolerated for SNe and stellar winds. Otherwise, the galaxy gets 
       destroyed at early epochs as a consequence of the energy released by 
       the star formation activity.
     \item In order not to overestimate the mean metallicity of the stars, in 
       the context of our models the infall time-scale has to be very short: 
       most of the gas must be available since the very beginning of the 
       computation to dilute the metals ejected by the first massive stars. We 
       set $\tau =$ 50 Myr for all models.
   \end{enumerate}


   \begin{figure}
     \begin{center}
     \includegraphics[width=\columnwidth]{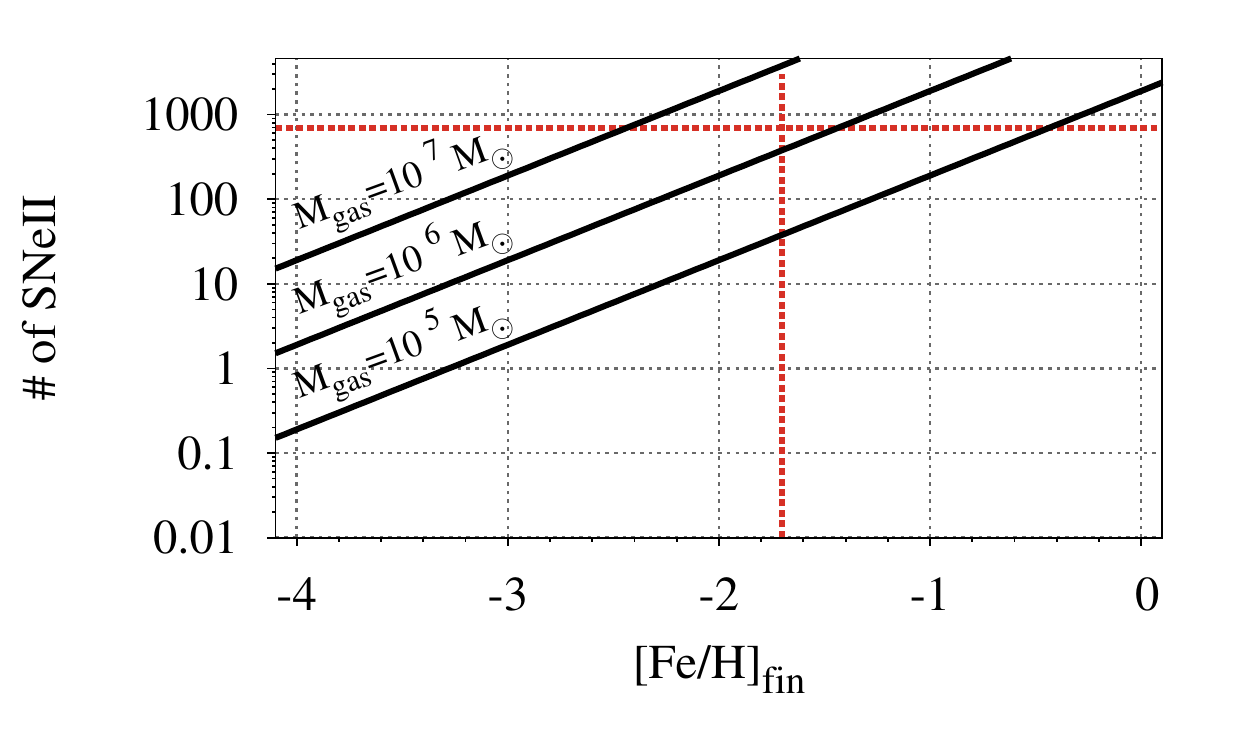}
     \caption{ Number of SNeII necessary to reach [Fe/H]$_{\mathrm{fin}}$, for 
       different gas cloud masses (diagonals) in the absence of gas 
       inflow/outflows. The horizontal dashed line marks the maximum number of 
       SNeII that are expected to have exploded in Bo\"otes~I, while the 
       vertical dashed line marks the maximum metallicity of Bo\"otes~I stars.}
     \label{fig:simple}
     \end{center}
   \end{figure}



   \begin{figure}
     \begin{center}
     \includegraphics[width=\columnwidth]{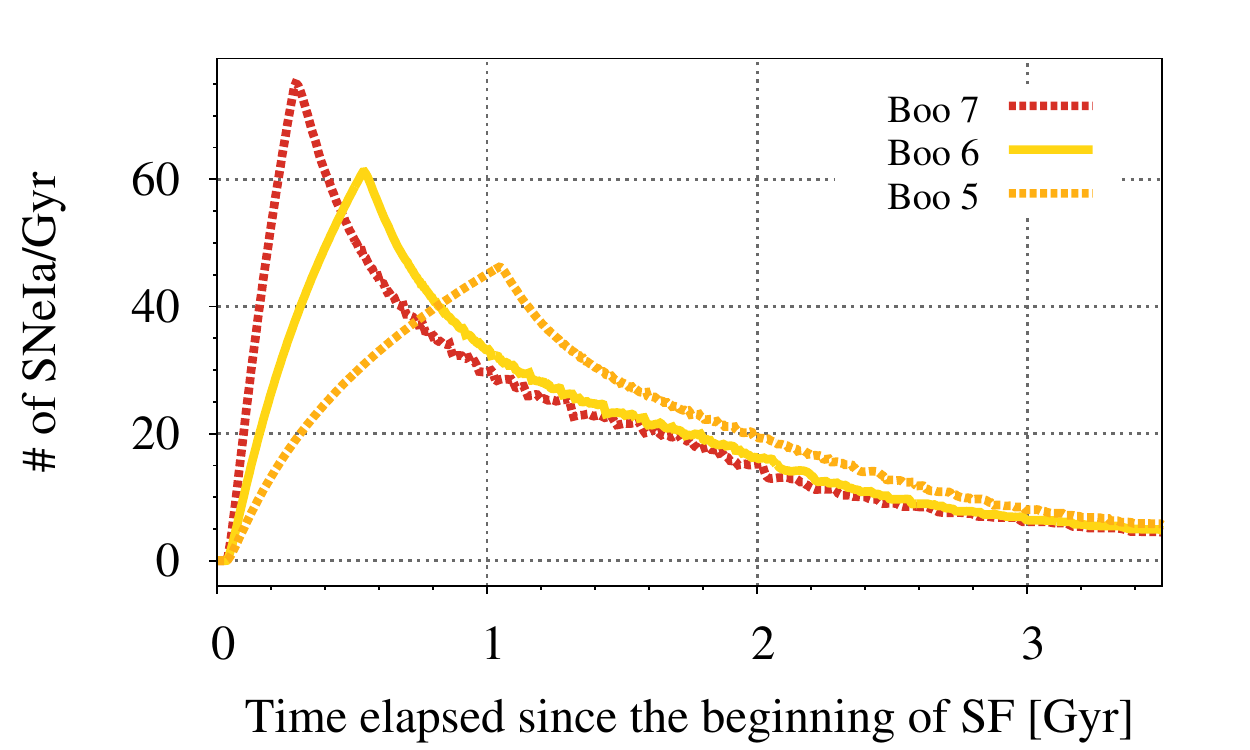}
     \caption{ SNIa rates (in number of events per Gyr) as functions of the 
       time elapsed since the beginning of the star formation predicted by 
       models~Boo\,{\sevensize 5}, Boo\,{\sevensize 6} and Boo\,{\sevensize 7}. 
       The duration of the star formation is different in the three cases, 
       namely it is 1, 0.5 and 0.25 Gyr, respectively.}
     \label{fig:SNeIa}
     \end{center}
   \end{figure}


   In Fig.~\ref{fig:mdfhom} we show the MDFs predicted by selected homogeneous 
   models for Bo\"otes~I (the MDFs of models~Boo\,{\sevensize 1}, 
   Boo\,{\sevensize 2} and Boo\,{\sevensize 3} overlap each other). The 
   theoretical predictions are compared to the observed distribution of stars 
   in the expanded sample of Lai et al. (2011). In Fig.~\ref{fig:mgfehom} the 
   run of [Mg/Fe] versus [Fe/H] is shown for the same models, in comparison 
   with high-resolution data from different authors (we exclude Boo-119, for 
   which Gilmore et al. 2013 measure [Mg/Fe]\,=\,1.04). All ratios are 
   normalized using solar reference values from Asplund et al. (2009). All 
   models assume one burst of star formation. While all the theoretical 
   [Mg/Fe]--[Fe/H] relations coarsely agree with the data, a satisfactory 
   agreement with the observed MDF is obtained only for a subset of models, 
   namely, the ones with the highest progenitor masses 
   (${\mathscr M}_{\mathrm{b}} \simeq$ 10$^7$~M$_\odot$; ${\mathscr M}_{\mathrm{DM}} =
   $ 6.5--11\,$\times$\,10$^7$~M$_\odot$; see next paragraph). In the framework 
   of the simple models considered here, the two stars with [Mg/Fe]~$\ge$ 0.5 
   at [Fe/H]~$\ge -$2\footnote{A high ratio of [Mg/Fe]\,=\,0.71$\pm$0.09 in 
     Boo-127 (Feltzing et al. 2009) is not confirmed by successive 
     investigations, that rather point to lower values, 
     [Mg/Fe]\,=\,0.17$\pm$0.18 (Gilmore et al. 2013) or 
     [Mg/Fe]\,=\,0.11$\pm$0.07 (Ishigaki et al. 2014).} can be explained only 
   as forming in a second, short-lasting --but extremely efficient-- starburst 
   involving  a minor fraction of unprocessed gas (see Figs.~\ref{fig:mdfhom2} 
   and \ref{fig:mgfehom2}; see also Table~\ref{tab:mod}, model~Boo\,{\sevensize 
     8}). Such an episode could be triggered by the interaction with the Milky 
   Way.

   The need for a substantially more massive progenitor can be easily 
   understood. Let us assume that each core-collapse SN produces 0.07~M$_\odot$ 
   of $^{56}$Ni (later decaying in $^{56}$Fe, Hamuy 2003) and that these ejecta 
   fully mix within the considered volume. In such a hypothesis, and in the 
   absence of inflow/outflow, $<$2~SNII explosions already suffice to increase 
   the metal content of a 10$^5$~M$_\odot$ gas cloud from [Fe/H]~$\simeq -$4 to 
   [Fe/H]~$\simeq -$3, while a metallicity [Fe/H]~$\simeq -$2 is reached after 
   19 such events. Since up to 700 SNeII are expected to have exploded in 
   Bo\"otes~I (assuming a canonical IMF), we end up with the request that some 
   10$^6$~M$_\odot$ of gas must have been present in order to dilute the freshly 
   produced metals and not to shift the stellar MDF towards values higher than 
   observed (see Fig.~\ref{fig:simple}). We also note that some SNeIa are 
   expected to have exploded in the system while it was still forming stars 
   (see Fig.~\ref{fig:SNeIa}). Since each SNIa releases 0.6--0.8~M$_\odot$ of Fe 
   (Iwamoto et al. 1999), even more diluting gas is needed. Although the above 
   discussion does not take into account the effect of large-scale outflows, 
   that could efficiently remove most of the metals from active star-forming 
   regions, nor the diluting effect of infall of unprocessed gas, it does give 
   a good idea of the global magnitudes of gas needed to produce a 
   Bo\"otes~I-like population.

   It is also worth stressing that if the IMF is significantly different from 
   what assumed here, our rough estimate changes. Weidner \& Kroupa (2005) 
   claim that steeper IMF slopes are to be expected for systems experiencing 
   low star formation rates, which would result in a number of core-collapse 
   SNe in Bo\"otes~I lower (or even significantly lower) than for a canonical 
   IMF.

   \subsubsection{Models in a cosmological context}


   \begin{figure*}
     \begin{center}
     \hspace{.2cm}
     \includegraphics[width=12cm]{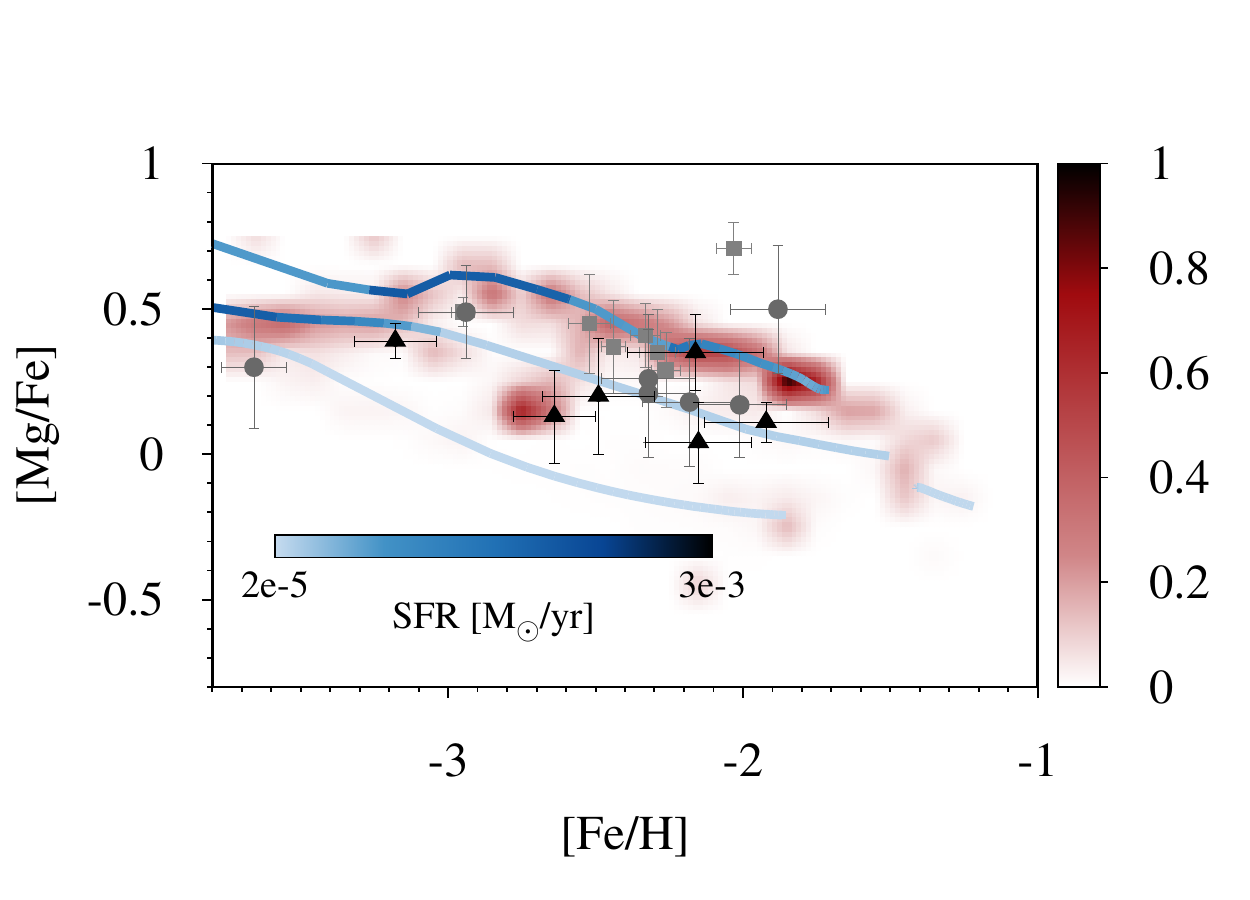}
     \caption{ [Mg/Fe] versus [Fe/H] for the models run in a full cosmological 
       context. The density map shows the distribution of long-lived stars for 
       eleven Bo\"otes~I candidates selected from the Starkenburg et al. (2013) 
       mock catalogue of Milky Way's satellites. The distribution is normalized 
       to its maximum value. The curves show the predictions of three 
       representative models, colour-coded according to their star formation 
       rates (legend at the bottom of the plot). Symbols with error bars are 
       high-resolution data for giant stars in Bo\"otes~I from Feltzing et al. 
       (2009; squares), Gilmore et al. (2013; circles) and Ishigaki et al. 
       (2014; triangles).}
     \label{fig:mgfecosmo}
     \end{center}
   \end{figure*}


   In classical chemical evolution models, a number of free parameters are 
   introduced in order to deal with poorly known physical processes such as gas 
   condensation, star formation and supernova feedback, and the interaction 
   with the environment is not taken into account. Semi-analytic models of 
   galaxy formation coupled with $N$-body (dark matter only) cosmological 
   simulations consider environmental effects to describe how gas (and stars) 
   get into galaxies, thus removing the need for a most uncertain 
   parameterization of the mass assembly histories of galaxies.
      

   \begin{figure*}
     \vspace{-1.1cm}
     \begin{tabular}{cc}
     \vspace{-1.1cm}
     \includegraphics[width=7cm]{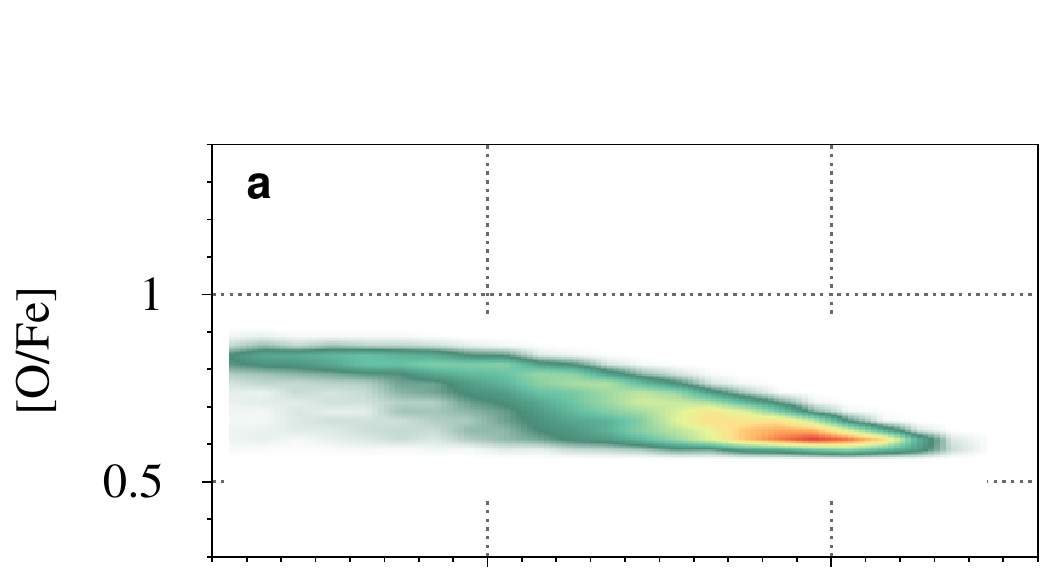} &
     \includegraphics[width=7cm]{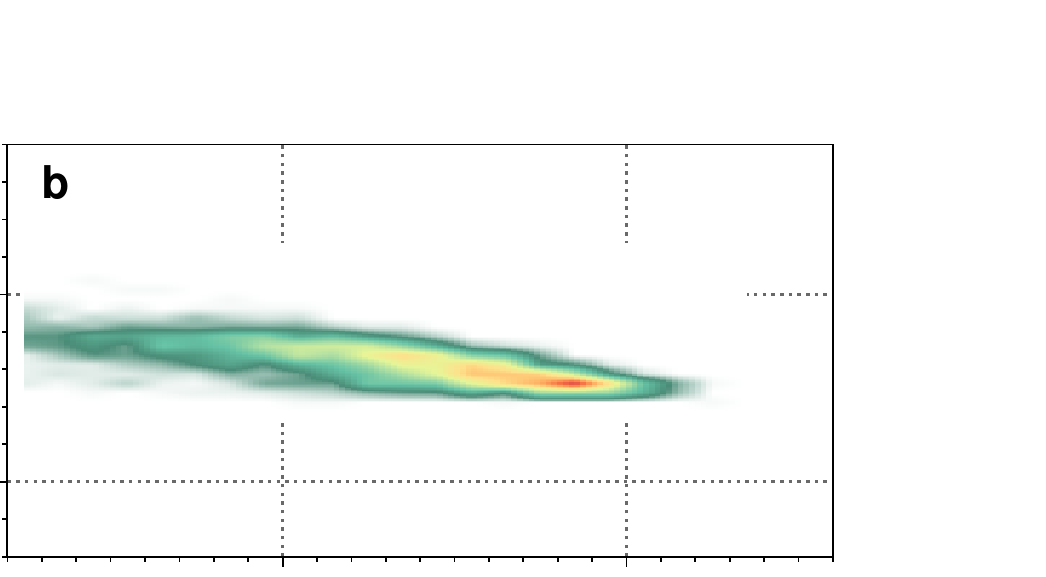} \\
     \vspace{-1.1cm}
     \includegraphics[width=7cm]{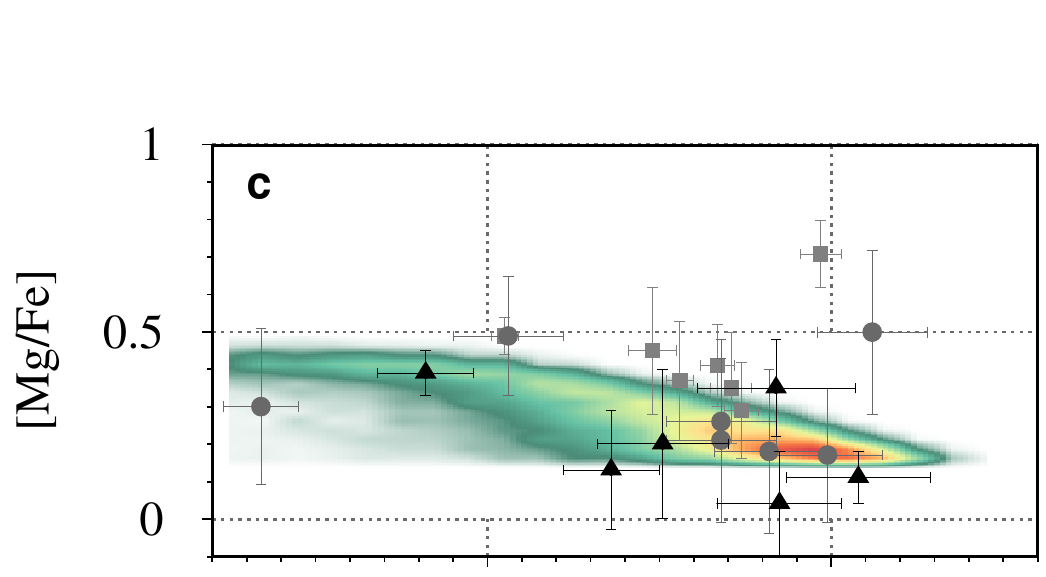} &
     \includegraphics[width=7cm]{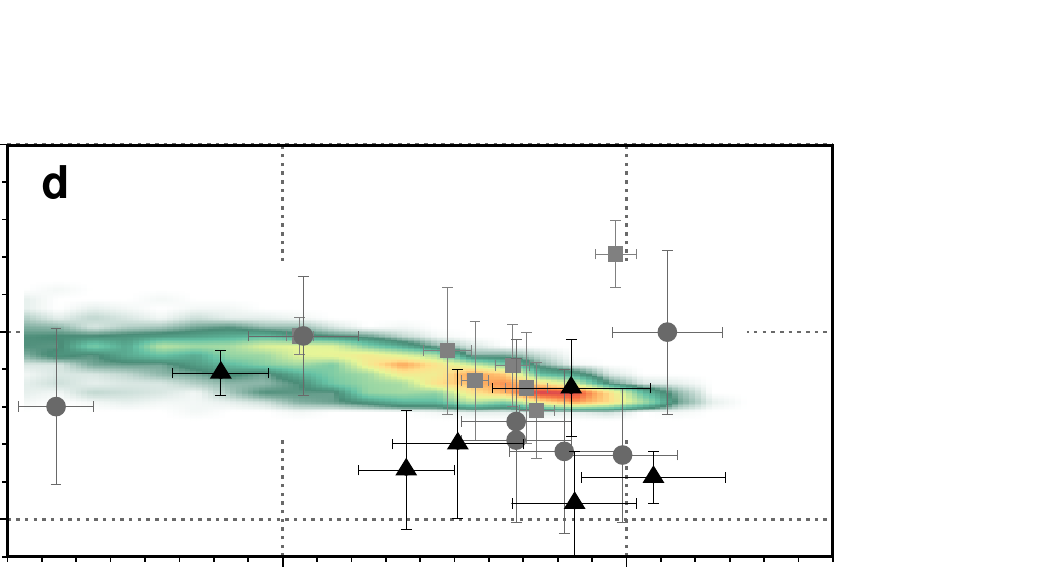} \\
     \vspace{-1.1cm}
     \includegraphics[width=7cm]{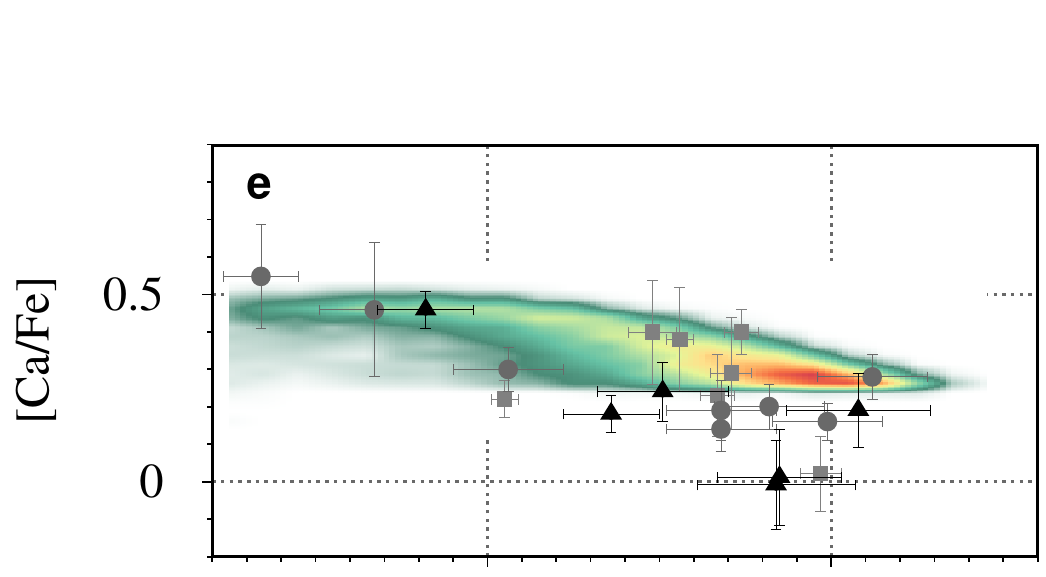} &
     \includegraphics[width=7cm]{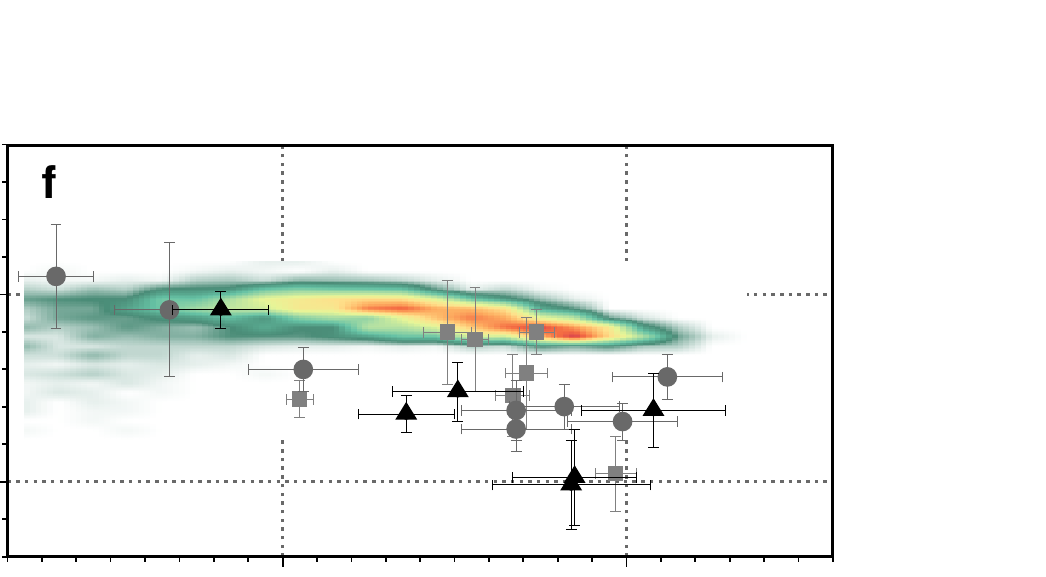} \\
     \vspace{-1.1cm}
     \includegraphics[width=7cm]{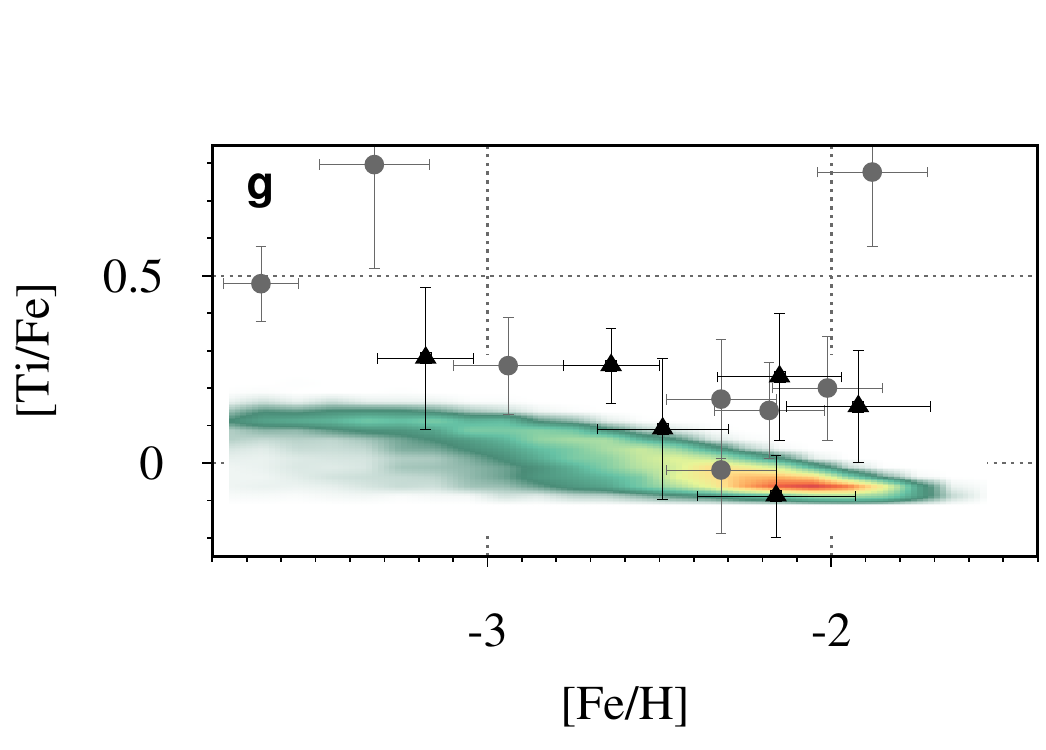} &
     \includegraphics[width=7cm]{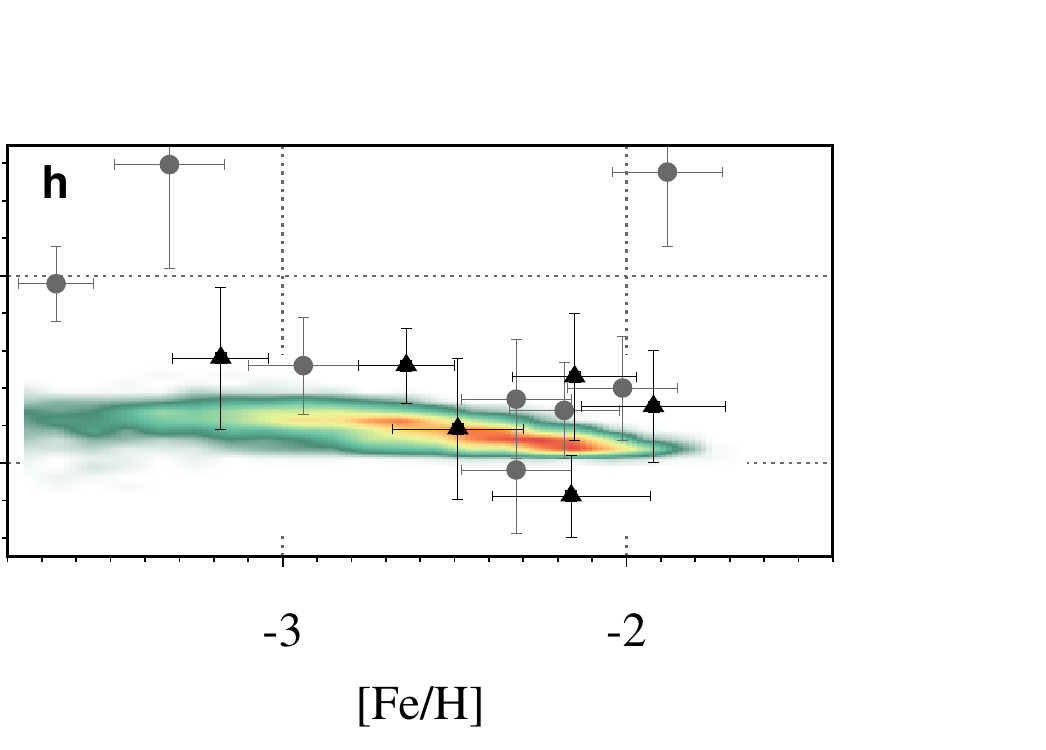} \\
     \multicolumn{2}{c}{\includegraphics[]{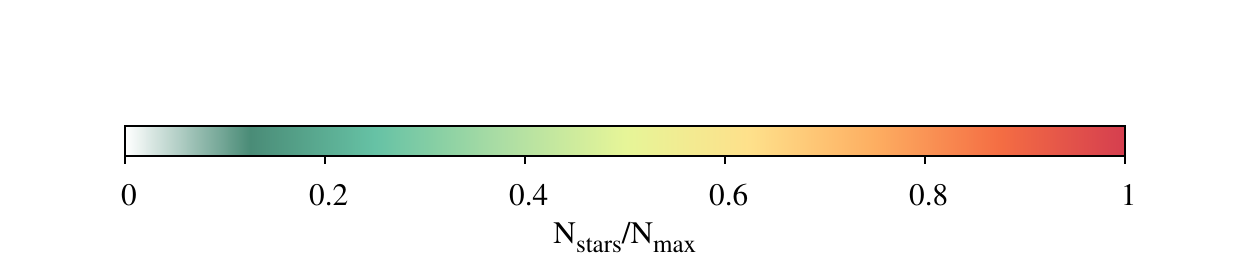} }
     \end{tabular}
     \caption{ [X/Fe] versus [Fe/H] for X~= O (panels a, b), Mg (panels c, d), 
       Ca (panels e, f) and Ti (panels g, h). The density maps show the 
       distributions of long-lived stars for models~Boo\,{\sevensize 5} (left 
       panels) and Boo\,{\sevensize 7} (right panels) when chemical 
       inhomogeneities are implemented following Leaman's (2012) empirical 
       relation between mean metallicity and metallicity spread. Each 
       distribution is normalized to its maximum value. Superimposed are 
       high-resolution data for giant stars in Bo\"otes~I (squares: Feltzing et 
       al. 2009; circles: Gilmore et al. 2013; triangles: Ishigaki et al. 
       2014).}
     \label{fig:inhalpha}
   \end{figure*}



   \begin{figure*}
     \vspace{-1.1cm}
     \begin{tabular}{cc}
     \vspace{-1.1cm}
     \includegraphics[width=7cm]{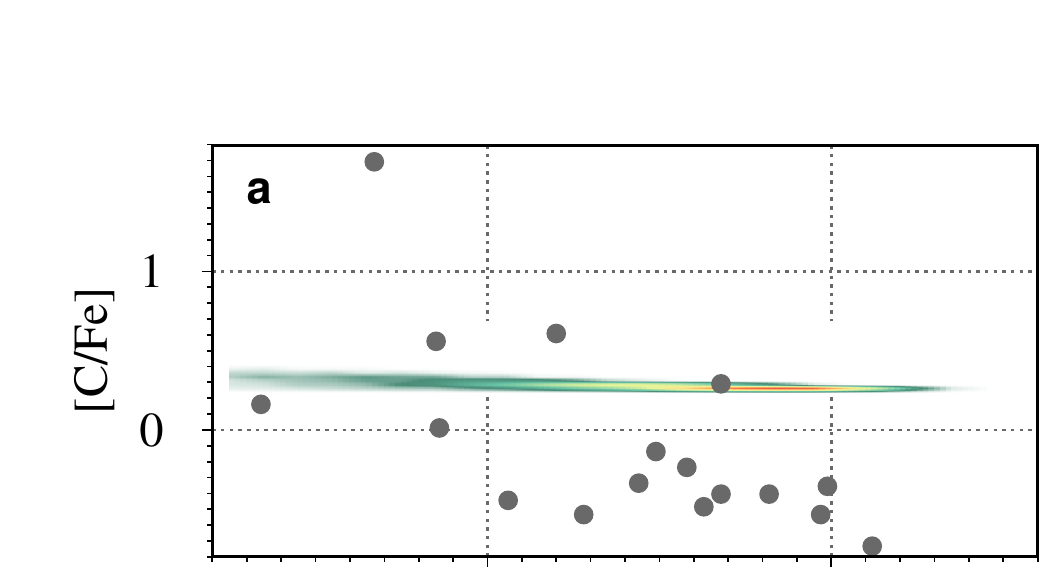} &
     \includegraphics[width=7cm]{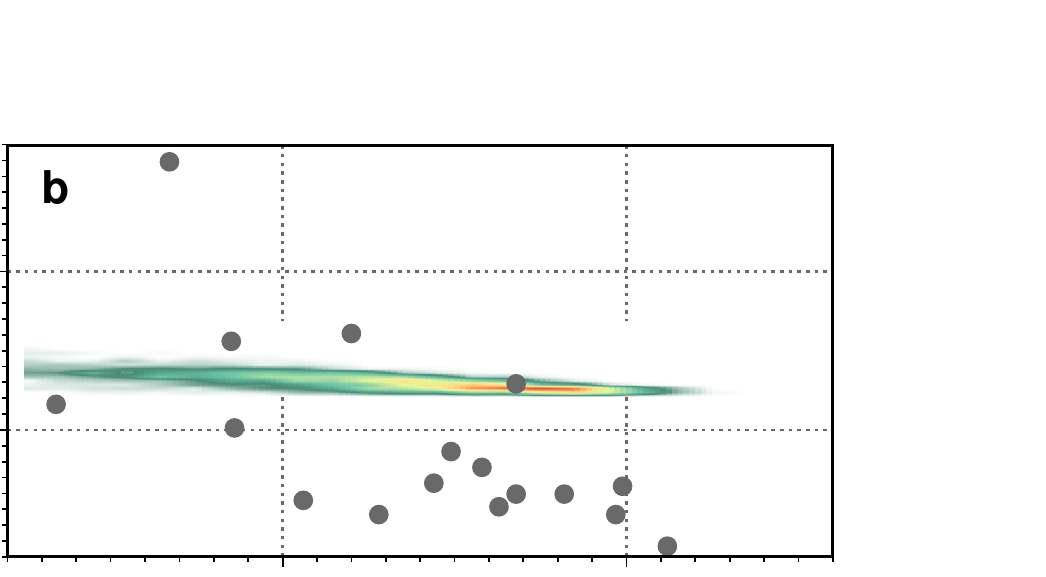} \\
     \vspace{-1.1cm}
     \includegraphics[width=7cm]{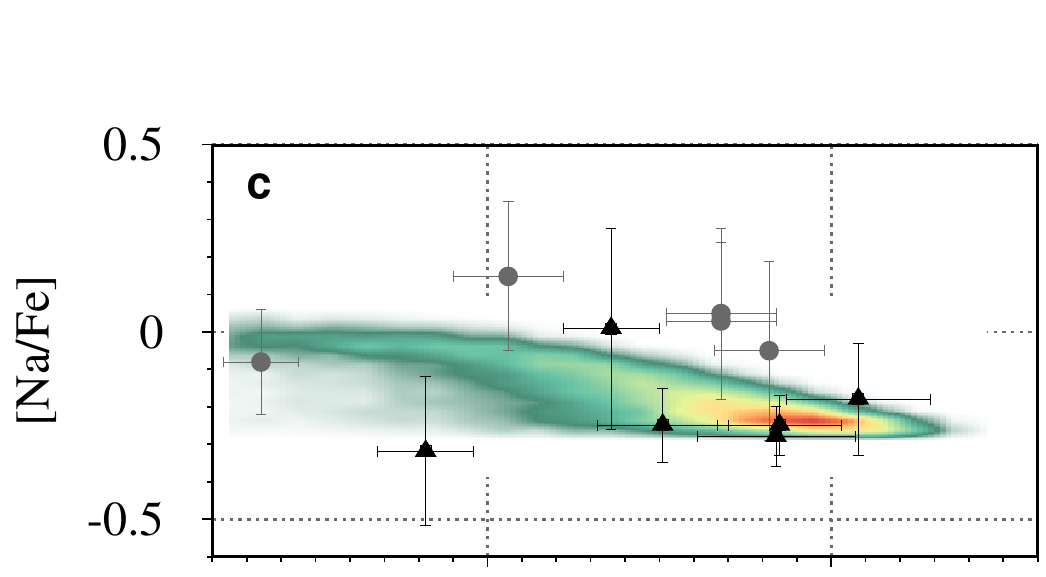} &
     \includegraphics[width=7cm]{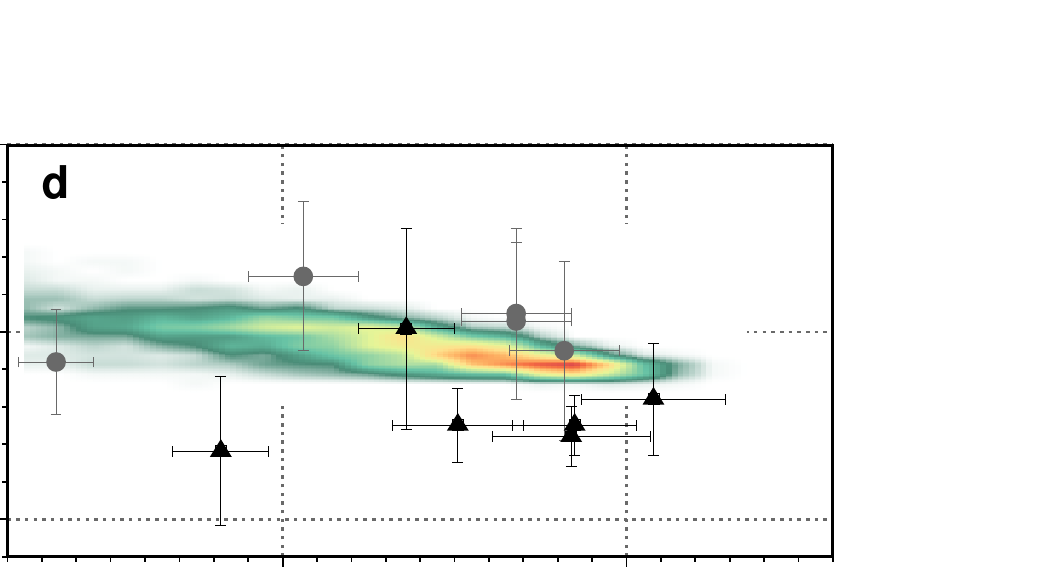} \\
     \vspace{-1.1cm}
     \includegraphics[width=7cm]{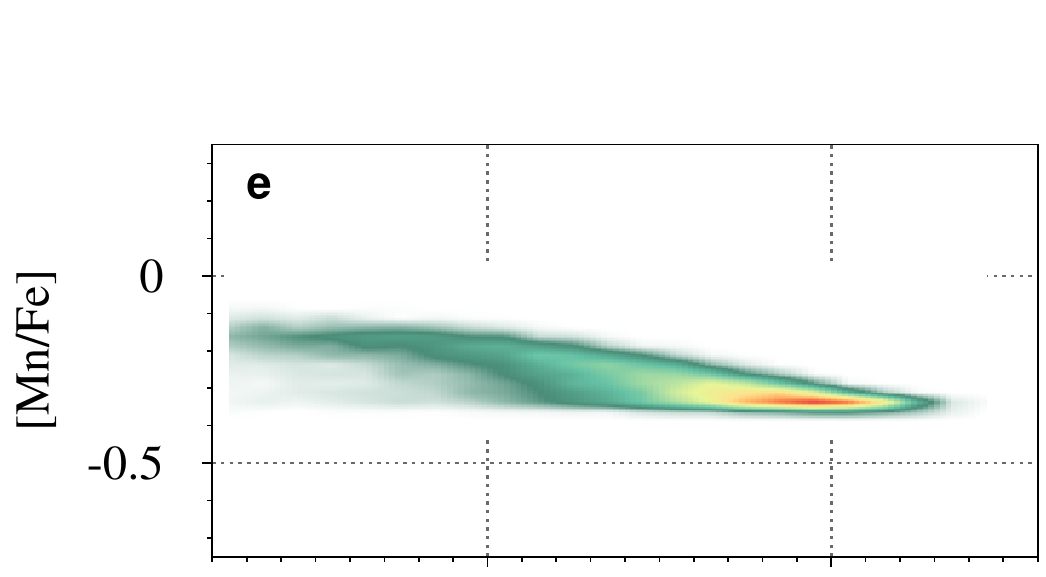} &
     \includegraphics[width=7cm]{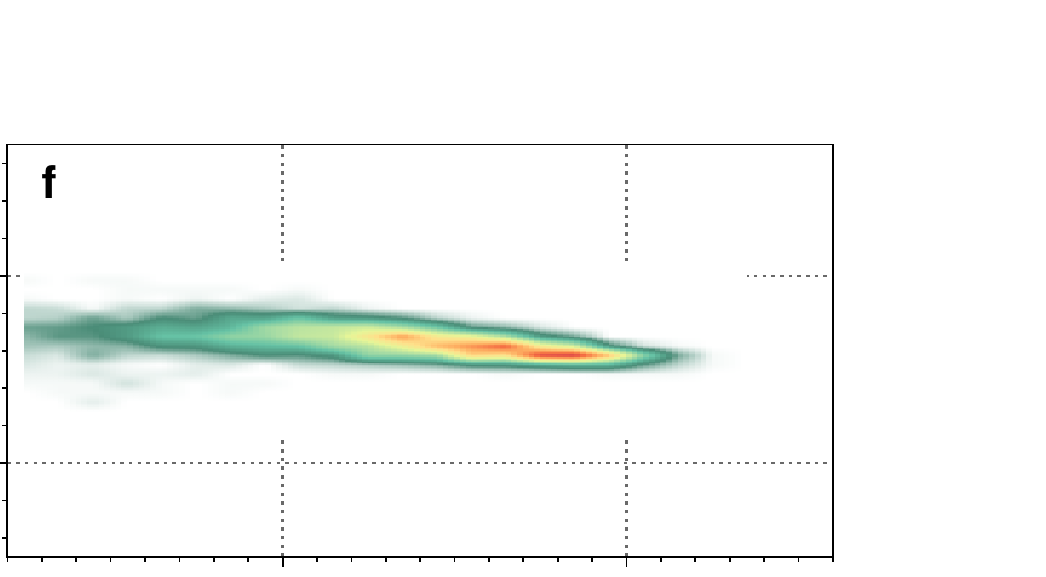} \\
     \vspace{-1.1cm}
     \includegraphics[width=7cm]{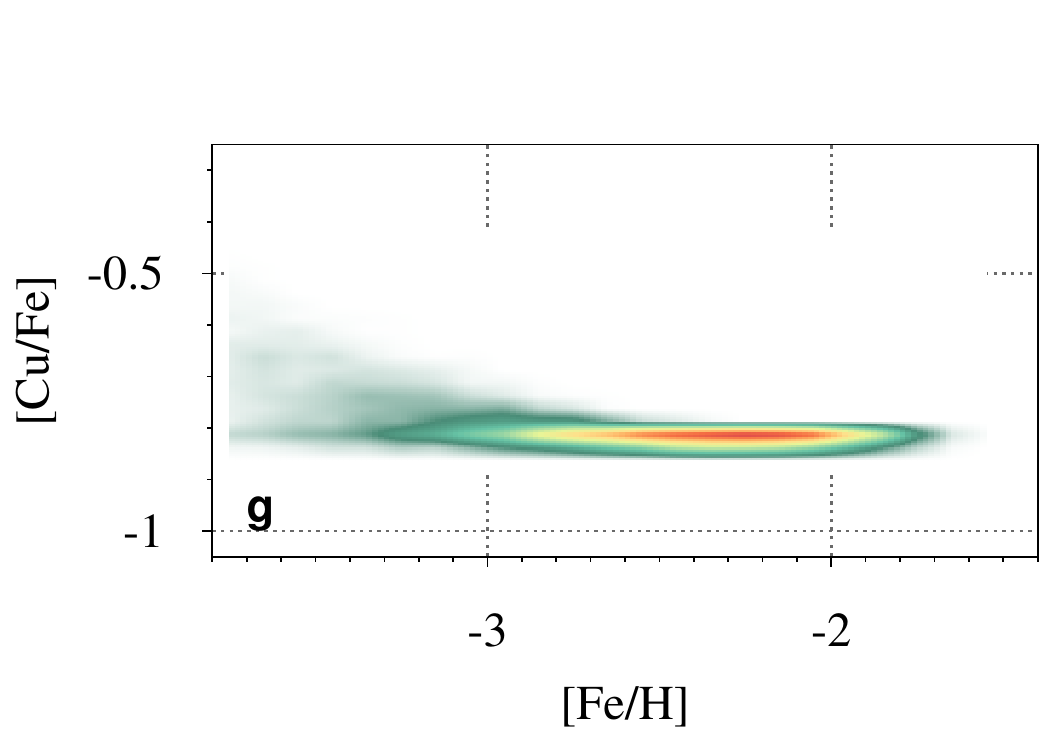} &
     \includegraphics[width=7cm]{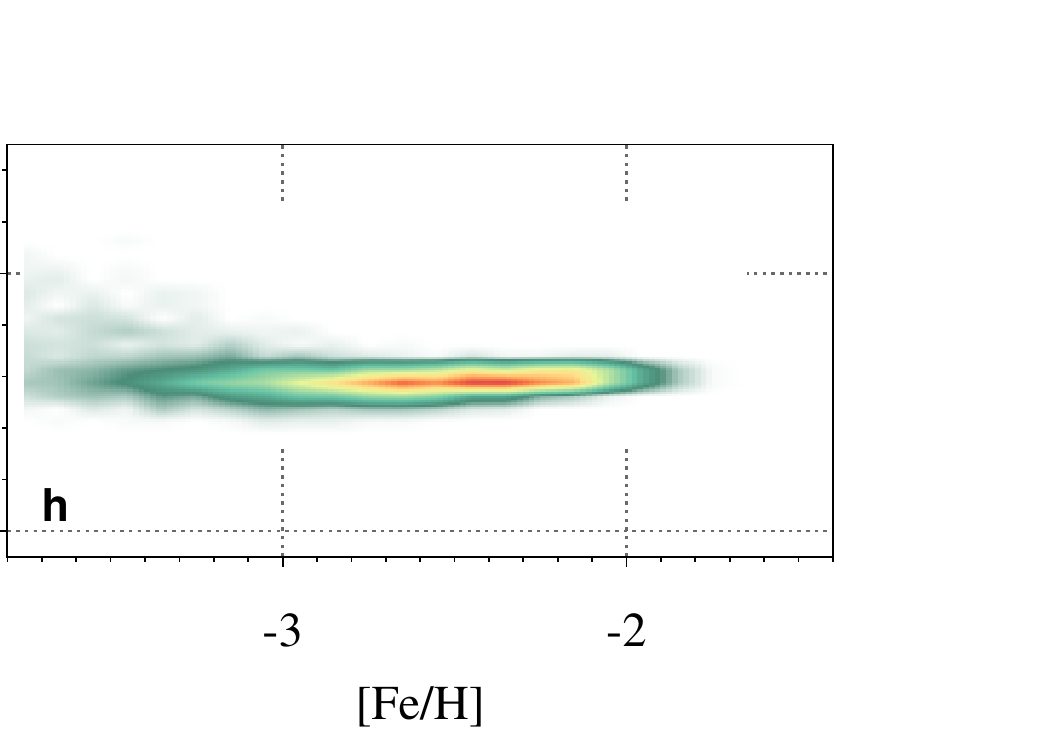} \\
     \multicolumn{2}{c}{\includegraphics[]{bar_rg-eps-converted-to.pdf} }
     \end{tabular}
     \caption{ Same as Fig.~\ref{fig:inhalpha} for C (panels a, b), Na (panels 
       c, d), Mn (panels e, f) and Cu (panels g, h). Carbon data from Norris et 
       al. (2010a) and Lai et al. (2011, one star).}
     \label{fig:inhother}
   \end{figure*}


   The (red) density map in Fig.~\ref{fig:mgfecosmo} summarizes the results we 
   get by post-processing for their detailed chemical properties eleven 
   Bo\"otes~I candidates selected from the mock galaxy catalogue of Starkenburg 
   et al. (2013, see Section~\ref{sec:cosmo}). The theoretical distribution, 
   that is normalized to its maximum value, shows where synthetic stars are 
   most likely found in the [Mg/Fe] versus [Fe/H] plot. Superimposed are the 
   relevant high-resolution abundance data (see Section~\ref{sec:data}). 
   Overall, the model predictions agree with the observations within the 
   errors, apart for a few very low-[Mg/Fe] stars that are predicted by the 
   models, but not observed. The Bo\"otes~I candidates are then inspected 
   one-by-one; the lines in Fig.~\ref{fig:mgfecosmo} show three representative 
   models. Colour-coding refers to the star formation rate --the darker the 
   curve, the higher the star formation rate. It is seen that the `blobs' 
   appearing in the density plot are not due to single star formation bursts 
   but, rather, to the coaddition of different models. Indeed, each model spans 
   almost the whole metallicity range of actual Bo\"otes~I stars. Some models 
   can be rejected, as they clearly underestimate  the [Mg/Fe] ratio at all 
   metallicities (see the lower curve in Fig.~\ref{fig:mgfecosmo}, 
   representative of this category of models), but the majority of them predict 
   [Mg/Fe] ratios in reasonable to good agreement with the observations (e.g., 
   middle and upper curves in Fig.~\ref{fig:mgfecosmo}). Looking back at the 
   star formation and mass assembly histories predicted by the semi-analytic 
   model, it is seen that:
   \begin{enumerate}
   \item Models at odd with the observations display large cold gas masses (of 
     the order of 10$^7$~M$_\odot$) and low star formation rates 
     ($<$0.001~M$_\odot$ yr$^{-1}$) at the epoch of chemical enrichment; as a 
     consequence, their stellar ejecta are strongly diluted (see 
     Fig.~\ref{fig:simple}) and the predicted MDFs peak at [Fe/H] values lower 
     than observed\footnote{Notice that, following Starkenburg et al. (2013) 
       and Li et al. (2010), in the models run in a fully cosmological setting 
       only 5 per cent of the stellar ejecta is added directly to the cold gas 
       component; the remainder is stored in a hot ejected component, from 
       which may or may not be re-accreted by the system.}. 
   \item A good fit to the observed properties is obtained when accretion 
     and/or loss of matter properly compensate for the star formation activity; 
     this delicate balance is achieved by a few models.
   \end{enumerate}

   \subsection{Inhomogeneous models}

   In Figs.~\ref{fig:inhalpha} and \ref{fig:inhother} we show the behaviour of 
   several abundance ratios as a function of [Fe/H] predicted by two of our 
   classical models, model~Boo\,{\sevensize 5} and model~Boo\,{\sevensize 7} 
   (left and right panels in each figure, respectively), when chemical 
   inhomogeneities are implemented following the empirical relation between 
   mean metallicity and intrinsic metallicity spread found for local dwarf 
   galaxies by Leaman (2012; see Section~\ref{sec:inho}). These two models have 
   been chosen since they predict the lowest (model~Boo\,{\sevensize 5}) and 
   highest (model~Boo\,{\sevensize 7}) [Mg/Fe] ratios at any given [Fe/H], 
   while their MDFs both agree reasonably well with the observational data.
   
   In model~Boo\,{\sevensize 5} the star formation is less efficient --and 
   lasts longer-- than in model~Boo\,{\sevensize 7}. Therefore, in 
   model~Boo\,{\sevensize 5} the ISM gets significantly enriched in iron by 
   SNeIa, at variance with model~Boo\,{\sevensize 7}. This results in lower 
   element-to-iron ratios for [Fe/H]~$> -$2.5 for all elements, in better 
   agreement with the bulk of the observations. A relatively long-lasting star 
   formation seems, hence, favoured in the frame of our models. Also, the 
   spread in the abundance ratios is reasonably well reproduced, with the 
   notable exception of carbon. It has been suggested (Gilmore et al. 2013, and 
   references therein) that this element might suffer a complex evolutionary 
   history, with two distinct enrichment channels active at very low 
   metallicities. This would explain both the CEMP-no and the carbon-normal 
   stars in Bo\"otes~I. A detailed study of carbon evolution should consider 
   the two different enrichment paths and is beyond the scope of the present 
   paper.

\begin{figure}
     \begin{center}
     \includegraphics[width=\columnwidth]{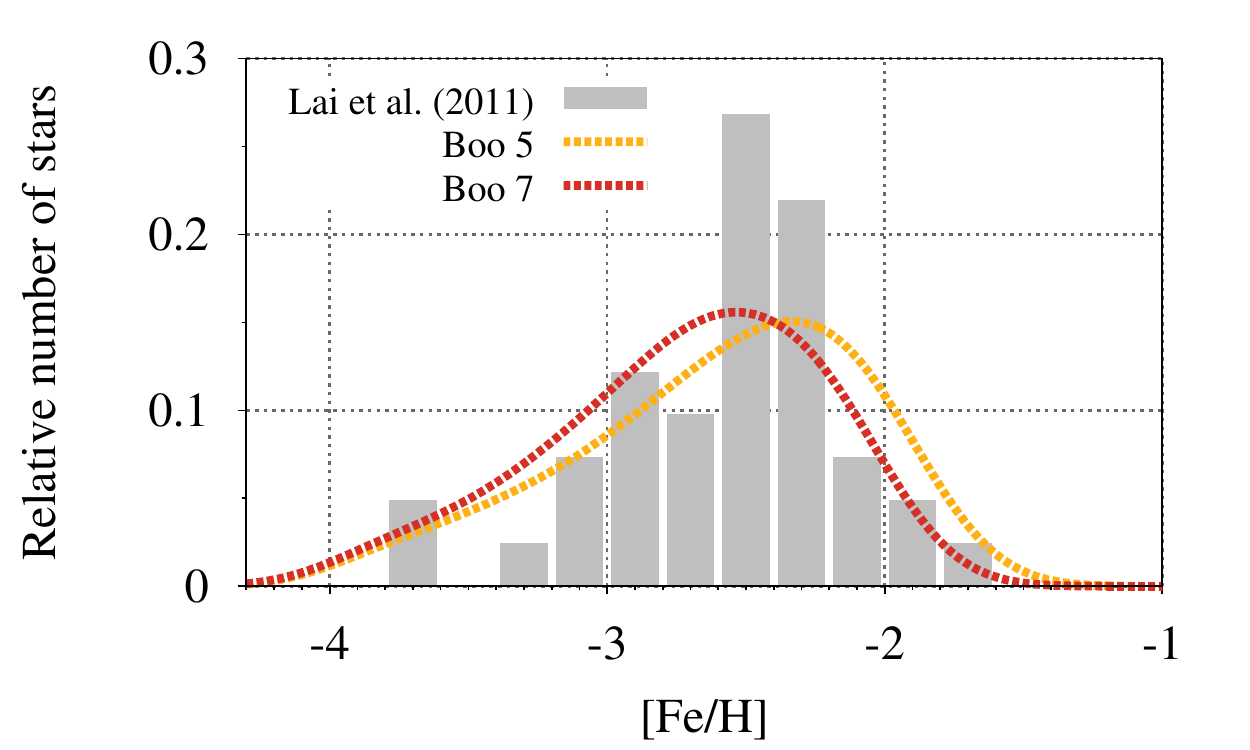}
     \caption{ Theoretical MDFs (dashed lines) predicted by 
       models~Boo\,{\sevensize 5} and Boo\,{\sevensize 7} when chemical 
       inhomogeneities are implemented, compared to the observed one 
       (histogram).}
     \label{fig:mdfinhom}
     \end{center}
   \end{figure}

   There are a few stars with anomalous abundances in one or more elements 
   other than carbon that can not be explained by our models:
   \begin{enumerate}
     \item Boo-119: this star has [Na/Fe]~= 0.73$\pm$0.23, [Mg/Fe]~= 
       1.04$\pm$0.22, [Ti/Fe]~= 0.80$\pm$0.28 (Gilmore et al. 2013). None of 
       these values can be explained by our models. Yet, [Ca/Fe] is normal in 
       this star ($\sim$0.45 dex).
     \item Boo-127: our models can not account for the high magnesium-to-iron 
       ratio, [Mg/Fe]~= 0.71$\pm$0.09, measured by Feltzing et al. (2009). 
       However, if [Mg/Fe]~= 0.17$\pm$0.18, as suggested by Gilmore et al. 
       (2013), or [Mg/Fe]~= 0.11$\pm$0.07 (Ishigaki et al. 2014) our models can 
       reproduce the data. The Na abundance measured by Ishigaki et al. (2014) 
       for this star, [Na/Fe]~= $-$0.18$\pm$0.15, is also well explained by the 
       models.
     \item Boo-41 and Boo-1137: both have [Ti/Fe]~$>$ 0.45 dex, well in excess 
       of the values predicted by our models, even when accounting for the 
       fact that the adopted Ti yields severely underestimate the trend of 
       [Ti/Fe] versus [Fe/H] for Galactic halo stars (see Romano et al. 2010, 
       their figure~22).
     \item Boo-117, Boo-127 and Boo-911: the nearly solar [Ca/Fe] ratios 
       reported by Felzing et al. (2009) for Boo-127 and Ishigaki et al. (2014) 
       for the other two objects can not be explained by our models. However, 
       higher ratios have been reported, that better match the model 
       predictions.
   \end{enumerate}
   Fig.~\ref{fig:mdfinhom} shows the MDFs predicted by models~Boo\,{\sevensize 
     5} and Boo\,{\sevensize 7} when chemical inhomogeneities are implemented 
   in the code (dashed lines). Adding the inhomogeneities leads to wider 
   theoretical MDFs (cfr. Fig.~\ref{fig:mdfhom}), that are still broadly 
   consistent with the observed distribution (grey histogram).
   
   We close this section with a note of caution. Combining data sets from 
   different authors, all coming with their different, methodology-dependent 
   systematics, may result in overestimating the intrinsic abundance 
   dispersions, if not even detecting spurious abundance spreads. The 
   homogeneous analysis of a statistically significant sample of Bo\"otes~I 
   stars is badly needed to shed light on important issues such as the 
   existence and significance of abundance spreads and the role of SNeIa in 
   enriching the ISM of Bo\"otes~I, one of the smallest known Milky Way 
   companions.

   \section{Summary and conclusions}

   The choice of Bo\"otes~I as a case of study was dictated by several reasons. 
   First, while it has a total luminosity nearly ten times smaller than the 
   faintest classical dwarfs, it is still one of the brightest among the 
   so-called UFDs ($L_V$\,=\,2.8$\times$10$^4$ M$_\odot$, also in the range of 
   globular clusters). The separation in two classes of the dwarf satellites 
   discovered before and after the advent of the Sloan Digital Sky Survey 
   (SDSS; York et al. 2000) is probably an arbitrary one (see, e.g., Belokurov 
   2013); still we are interested to the chemical evolution of stellar systems 
   that, at face value, have a total stellar mass far too low to retain the SN 
   ejecta (say, lower than 10$^6$ M$_\odot$). On the other hand, the smaller the 
   total stellar mass of a system, the lower the total number of stars, and, in 
   particular, of stars suitable for chemical analysis (i.e. red giant branch 
   stars, in the range of distances of dwarf satellites of the Milky Way). In 
   this framework Bo\"otes~I appears as a good trade-off system, with the 
   additional advantage of being relatively nearby. Probably, because of the 
   above characteristics it is also among the best studied UFDs, in particular 
   from the spectroscopic point of view (see Section~\ref{sec:data}). Hence, 
   relatively abundant observational constraints are available for comparison 
   with chemical evolution models. Finally, the galaxy is completely dominated 
   by very old stars with a small spread in age (de Jong et al. 2008; Hughes et 
   al. 2008)), hence the star formation history is quite simple.


   We have run and compared different chemical evolution models for Bo\"otes~I  
   --classical models versus models run in a fully cosmological setting, as 
   well as homogeneous versus inohomogeneous models-- and compared the model 
   predictions with the available observations. 
   \begin{enumerate}
   \item As for the classical models, we suggest, in agreement with Vincenzo et 
     al. (2014), that Bo\"otes~I must have formed through accretion of 
     relatively large amounts of gas (${\mathscr M}_{\mathrm{b}} 
     \simeq$10$^7$~M$_\odot$) on very short time scales ($\tau$~=~50~Myr) and 
     converted into stars less than 1 per cent of its baryons (see also 
     Salvadori \& Ferrara 2009). These conditions have to be met in order not 
     to overestimate the metal content of Bo\"otes~I stars. In the case of the 
     cosmologically-motivated models, lower amounts of diluting gas are needed, 
     because most of the stellar ejecta is stored in a hot ejected component, 
     rather than being mixed directly with the neutral ISM.
   \item At variance with Vincenzo et al. (2014), we do not find a clear-cut 
     evidence that the gas left over from the star formation process can be 
     expelled from the galaxy through large-scale outflows; rather, in the  
     classical approach we are left with huge amounts of gas and have to turn 
     to the cosmologically-motivated models to find that the residual gas is 
     most likely stripped by the interaction with the Milky Way. 
   \item Though some of the cosmologically-motivated models predict lower than 
     observed [$\alpha$/Fe] ratios in Bo\"otes~I, the majority of them provide 
     reasonable to good fit to the available data. Looking back at the star 
     formation and mass assembly histories predicted by the semi-analytical 
     model, we find that a delicate balance of mass loss, mass accretion and 
     star formation is needed for the model predictions to agree with the 
     observations. 
   \item Chemical inhomogeneities are implemented in our code following the 
     empirical relation between mean metallicity and metallicity spread found 
     by Leaman (2012) for dwarf galaxies in the Local Group. In this framework, 
     our capability to reproduce the observed spread in abundance ratios is 
     directly linked to the extent of the variation of the relative yields with 
     metallicity that is expected owing to the adopted stellar nucleosynthesis 
     prescriptions (see Appendix~A). 
   \item Full (three-dimensional) hydrodynamical simulations including stellar 
     feedback and chemical enrichment are needed in order to obtain better 
     insights on issues such as the development of outflows and the 
     establishment of inhomogeneities in Bo\"otes~I; we have recently embarked 
     on this kind of computations. 
   \end{enumerate}
   As a final remark, it is worth stressing that at present high-resolution 
   spectroscopic observations have been obtained only for a small sample of 
   Bo\"otes~I stars. This severely hampers our capability of drawing firm 
   conclusions on issues such as the impact of SNeIa in enriching the ISM of 
   Bo\"otes~I, or the significance of inhomogeneities.

   \section*{Acknowledgements}

   The authors are indebted to the Virgo Consortium, which was responsible for 
   designing and running the halo simulations of the Aquarius Project. 
   Additionally, they are grateful to Gabriella De Lucia, Amina Helmi and 
   Yang-Shyang Li for their role in developing the semi-analytic model of 
   galaxy formation used in this paper. DR thanks Francesca Matteucci for 
   countless discussions on chemical evolution and wise advice. Finally, we 
   wish to thank the anonymous referee for comments that improved the 
   presentation of the paper. DR and ES are indebted to the International Space 
   Science Institute (ISSI), Bern, Switzerland, for supporting and funding the 
   international team ``First stars in dwarf galaxies''. MB and DR acknowledge 
   financial support from PRIN~MIUR~2010--2011, project ``The Chemical and 
   Dynamical Evolution of the Milky Way and Local Group Galaxies'', 
   prot.~2010LY5N2T. DR acknowledges support from PRIN~INAF~2010, project 
   ``Looking for the elusive building blocks of the Milky Way and Andromeda 
   Halos''. RL acknowledges financial support to the DAGAL network from the 
   People Programme (Marie Curie Actions) of the European Union's Seventh 
   Framework Programme FP7/2007- 2013/ under REA grant agreement number 
   PITN-GA-2011-289313. ES gratefully acknowledges the Canadian Institute for 
   Advanced Research (CIfAR) Global Scholar Academy for support.


   \begin{figure*}   
     \vspace{-1.1cm}
     \begin{tabular}{cc}
     \vspace{-1.1cm}
     \includegraphics[width=8.4cm]{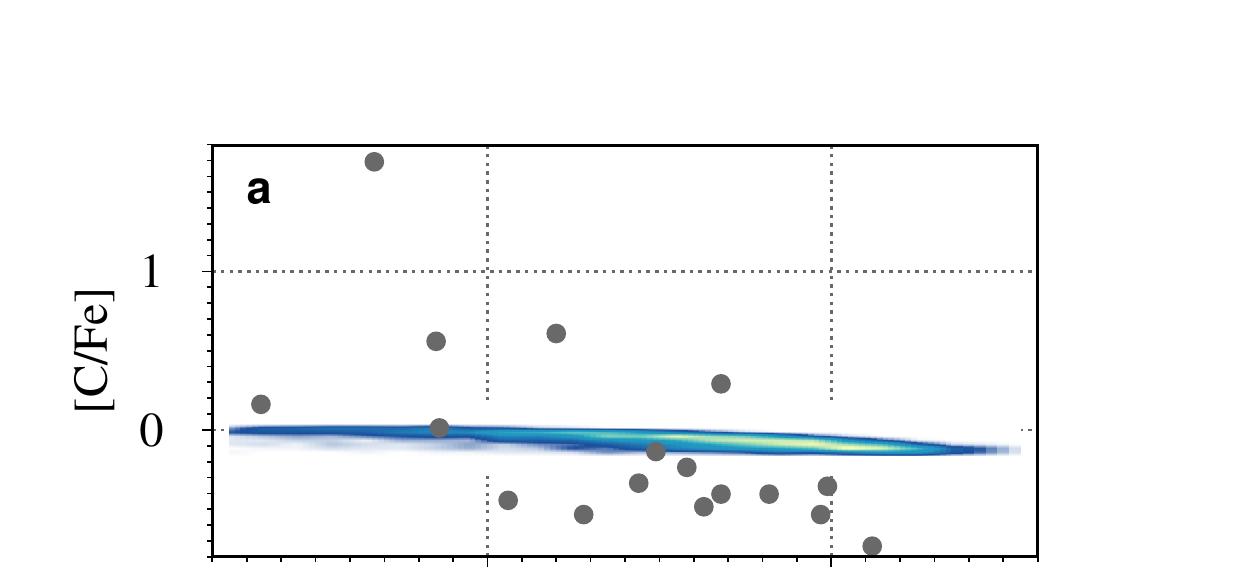} &
     \hspace{-1.1cm}
     \includegraphics[width=8.4cm]{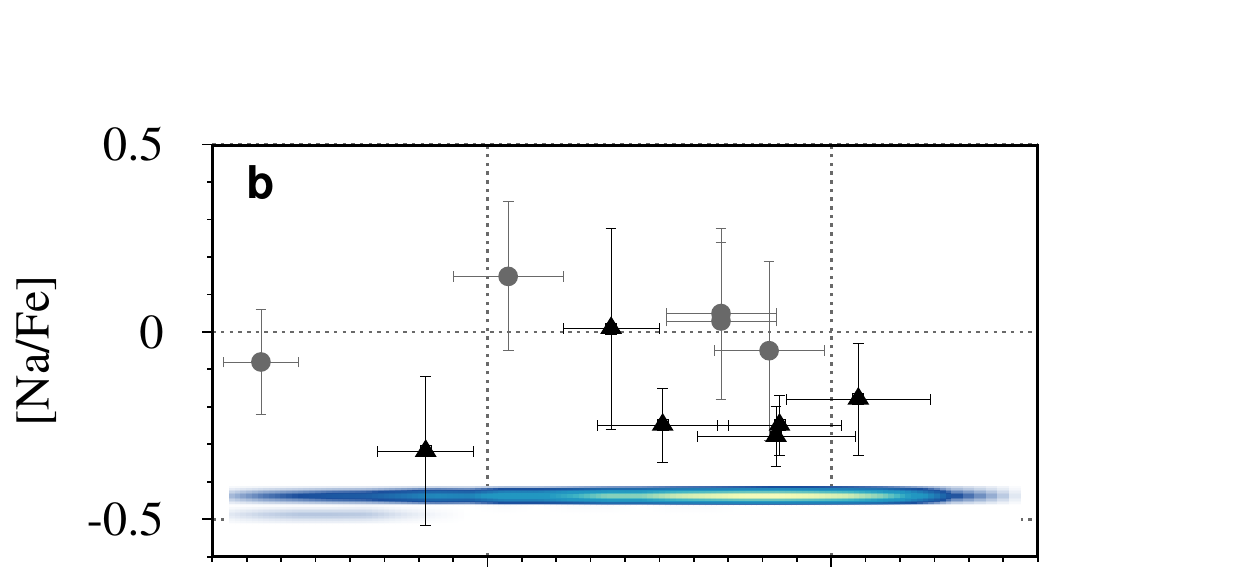} \\
     \vspace{-1.1cm}
     \includegraphics[width=8.4cm]{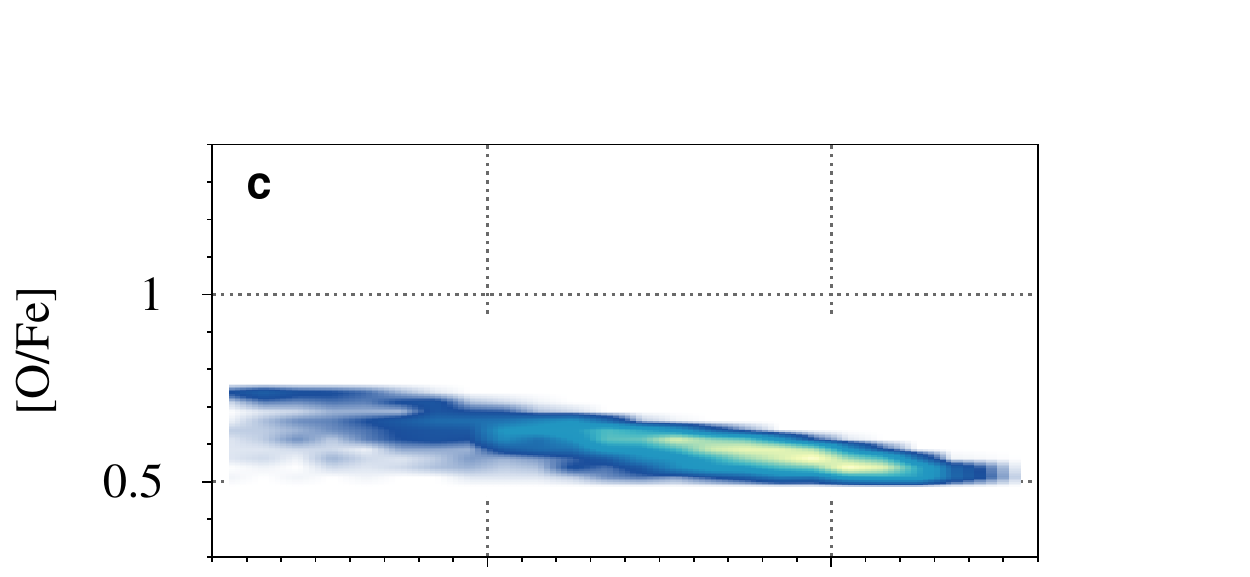} &
     \hspace{-1.1cm}
     \includegraphics[width=8.4cm]{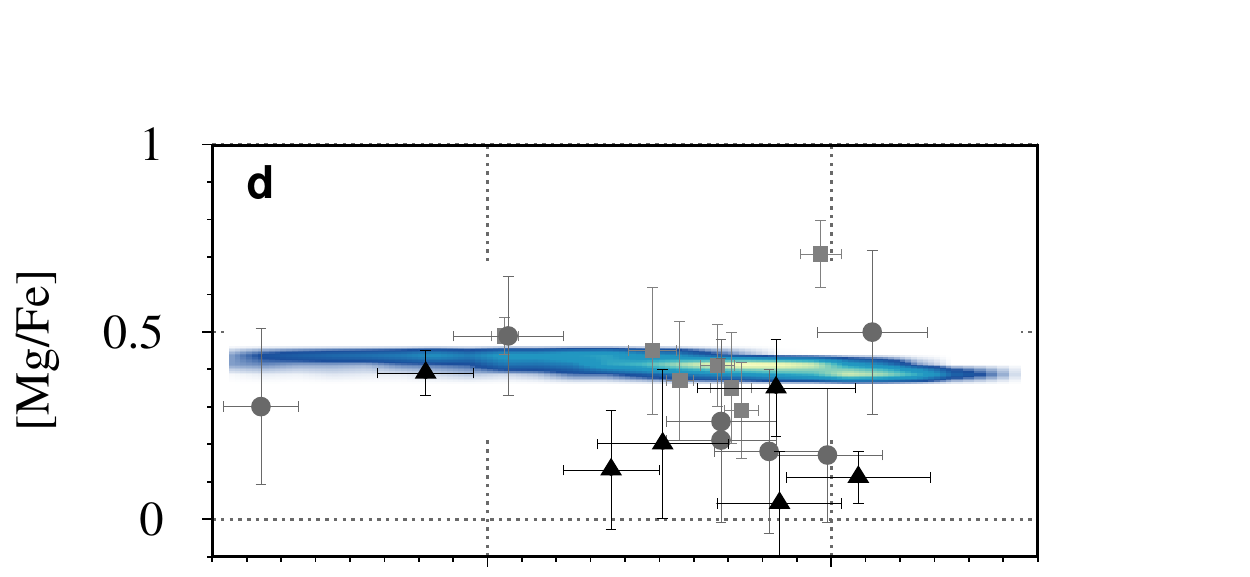} \\
     \vspace{-1.1cm}
     \includegraphics[width=8.4cm]{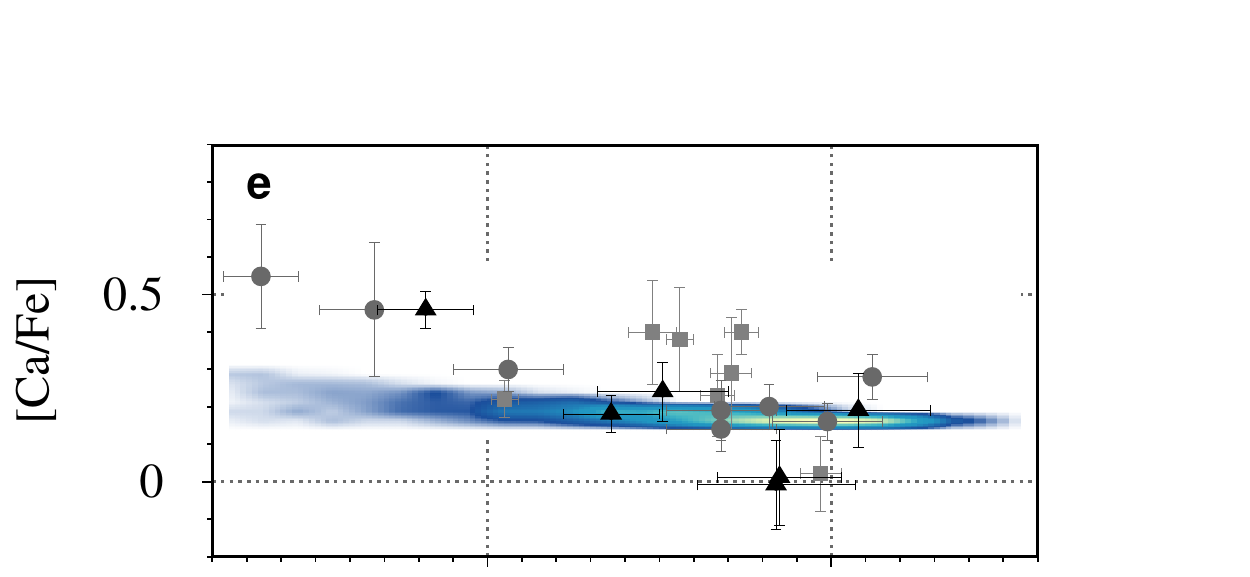} &
     \hspace{-1.1cm}
     \includegraphics[width=8.4cm]{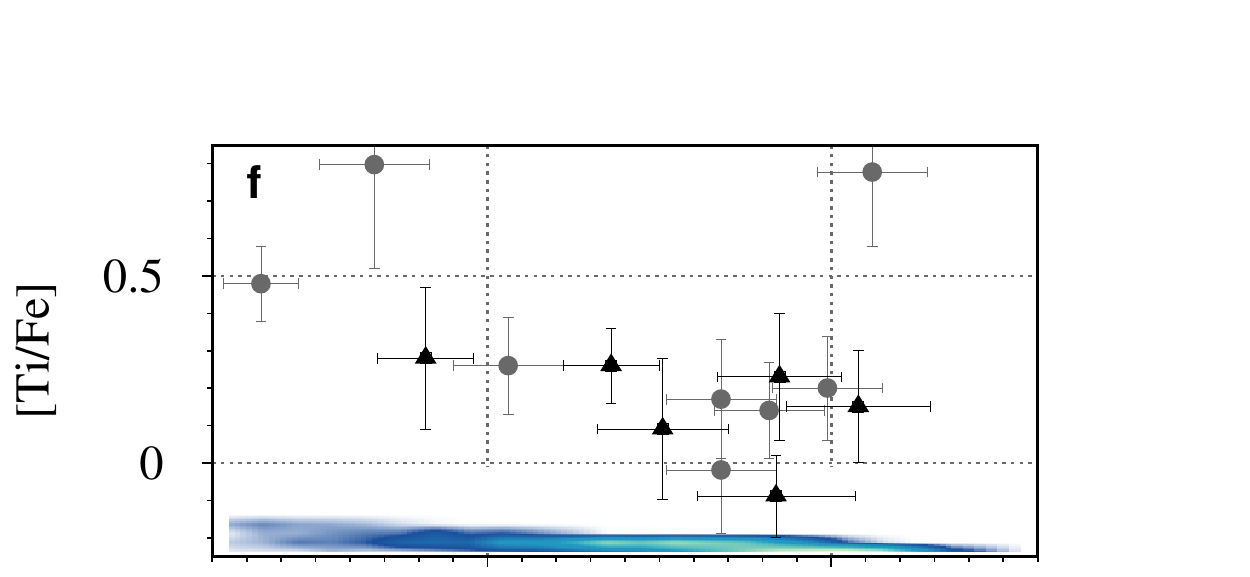} \\
     \vspace{-1.1cm}
     \includegraphics[width=8.4cm]{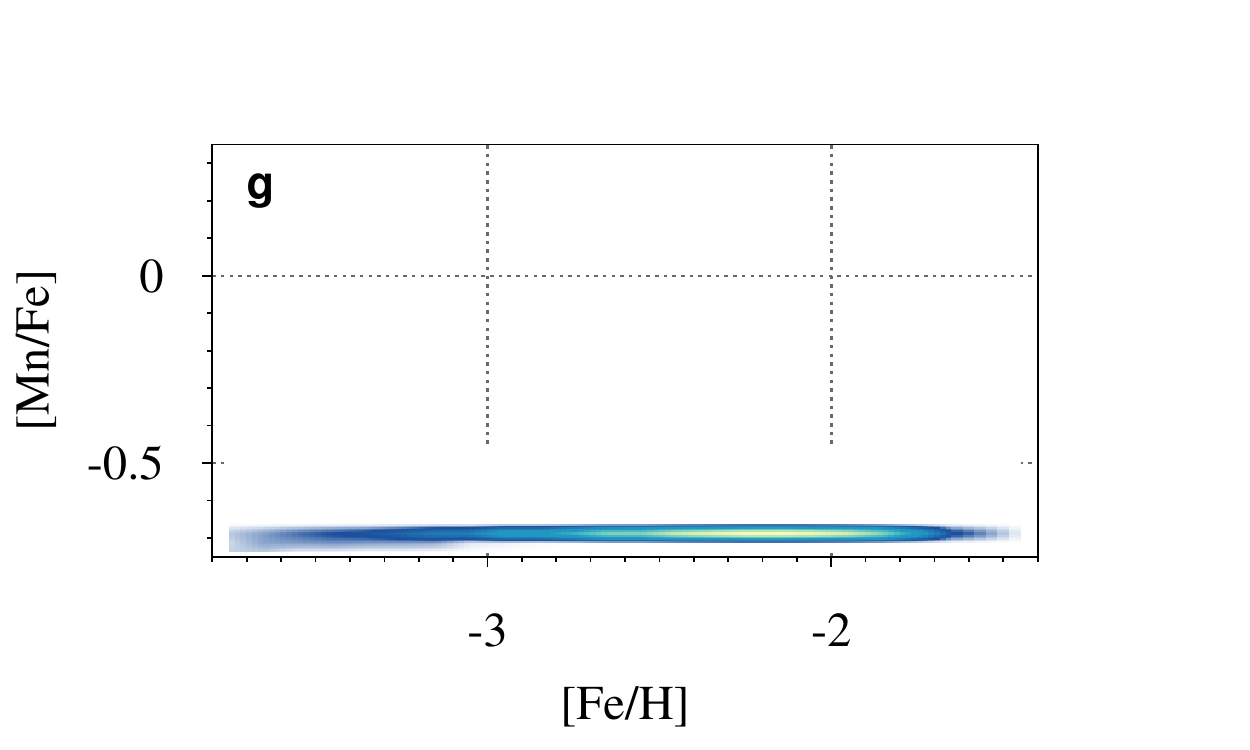} &
     \hspace{-1.1cm}
     \includegraphics[width=8.4cm]{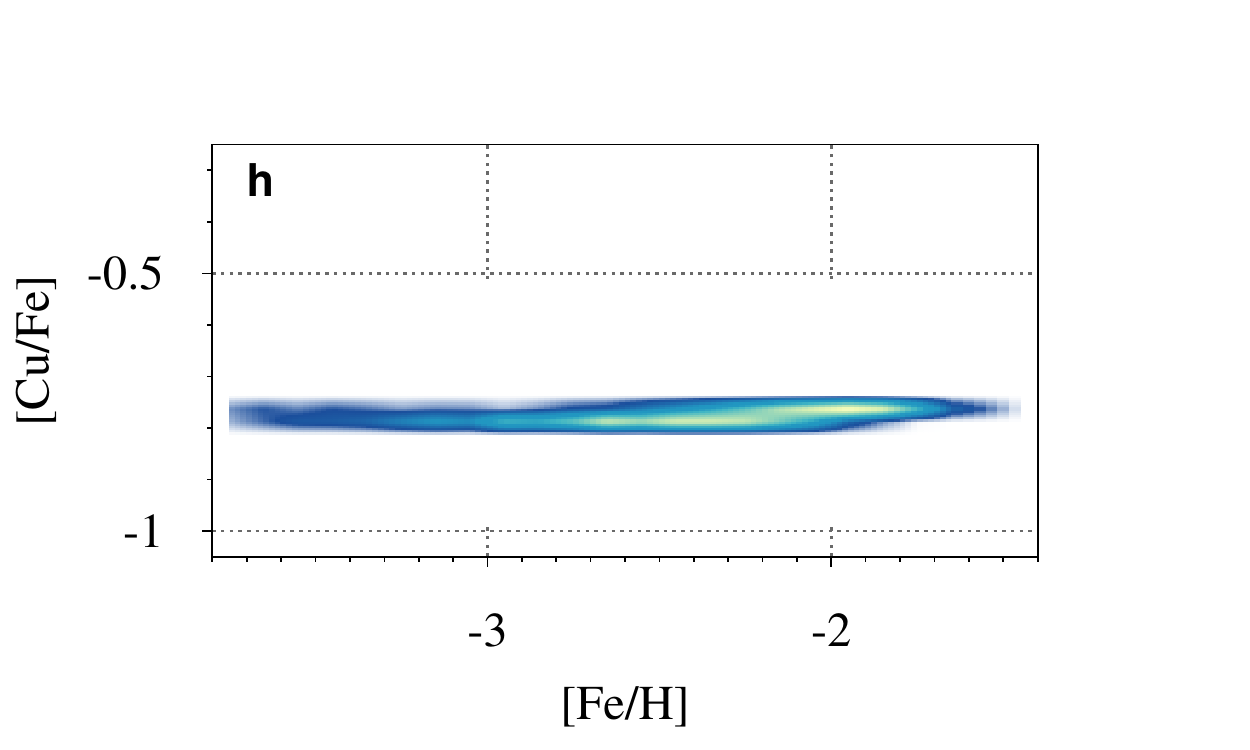} \\
     \multicolumn{2}{c}{\includegraphics[]{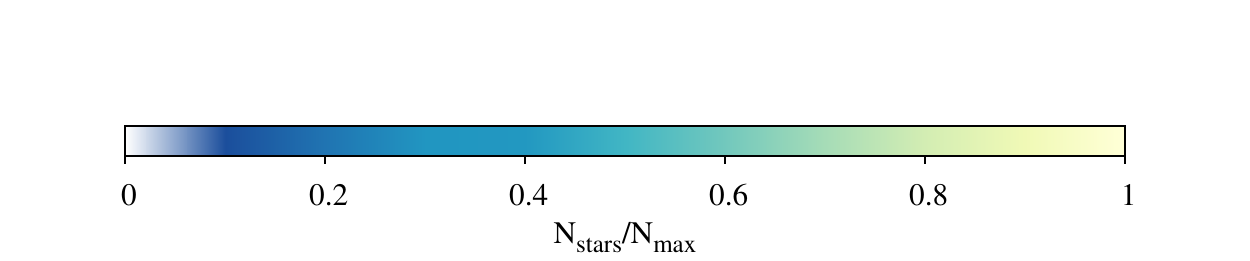} }
     \end{tabular}
     \caption{ [X/Fe]--[Fe/H] relations in Bo\"otes~I. The contours show the 
       frequency distribution of long-lived stars in the simulated galaxy, 
       where yellow is for the highest frequency and blue for the lowest one. 
       High-resolution data for giant stars are taken from Feltzing et al. 
       (2009; squares), Gilmore et al. (2013; circles) and Ishigaki et al. 
       (2014; triangles).}
     \label{fig:yields}
   \end{figure*}

\section*{Appendix A: The impact of the choice of the stellar yields}

In Fig.~\ref{fig:yields} we show the predictions of model~Boo\,{\sevensize 7} 
with chemical inhomogeneities implemented as described in Sect.~\ref{sec:inho}, 
for a different choice of the stellar yields, namely, Karakas (2010) for single 
low- and intermediate-mass stars and Kobayashi et al. (2006) for single massive 
stars exploding as SNeII (see Sect.~\ref{sec:chfdb}). When comparing the 
results shown in Fig.~\ref{fig:yields} with the corresponding ones relevant to 
our standard yield choice, displayed in the right-hand panels of 
Figs.~\ref{fig:inhalpha} and \ref{fig:inhother}, it is immediately seen that 
with the new nucleosynthesis prescriptions:
\begin{enumerate}
\item the bulk of the synthetic stars has higher metallicities;
\item lower [X/Fe] ratios are predicted at all [Fe/H] for C, Na, O, Ca, Ti, Mn, 
  and Cu, while the predicted [Mg/Fe] ratio is slightly higher for [Fe/H]~$\ge 
  -$2 dex;
\item a much smaller dispersion is expected for all the abundance ratios, with 
  the notable exception of [O/Fe].
\end{enumerate}
The first two points simply reflect the fact that lower/higher amounts of each 
chemical element are restored to the ISM by dying stars according to the 
different nucleosynthesis studies analysed here. The third issue, instead, has 
to do with the variation of the relative yields with metallicity, that for 
Kobayashi et al. (2006) is smaller than for Woosley \& Weaver (1995), at least 
for the metallicity range probed by this study.

     \label{lastpage}


\begin{thebibliography}{90}
\bibitem{}
Asplund M., Grevesse N., Sauval A. J., Scott P., 2009, ARA\&A, 47, 481
\bibitem{}
Bailin J., Ford A., 2007, MNRAS, 375, L41
\bibitem{}
Battaglia G., et al., 2006, A\&A, 459, 423
\bibitem{}
Belokurov V., 2013, New Astron. Rev., 57, 100
\bibitem{}
Belokurov V., et al., 2006, ApJ, 647, L111
\bibitem{}
Bradamante F., Matteucci F., D'Ercole A., 1998, A\&A, 337, 338
\bibitem{}
Carigi L., Hernandez X., 2008, MNRAS, 390, 582
\bibitem{}
Carretta E., Bragaglia A., Gratton R., D'Orazi V., Lucatello S., 2009, A\&A, 
508, 695
\bibitem{}
Cescutti G., 2008, A\&A, 481, 691
\bibitem{}
Conselice C. J., 2012, preprint (arXiv:1212.5641)
\bibitem{}
Cooper A. P., et al., 2010, MNRAS, 406, 744
\bibitem{}
Croton D. J., Springel V., White S. D. M., De Lucia G., Frenk C. S., Gao L., 
Jenkins A., Kauffmann G., Navarro J. F., Yoshida N., 2006, MNRAS, 365, 11
\bibitem{}
de Jong J. T. A., Rix H.-W., Martin N. F., Zucker D. B., Dolphin A. E., Bell
E. F., Belokurov V., Evans N. W., 2008, AJ, 135, 1361
\bibitem{}
De Lucia G., Blaizot J., 2007, MNRAS, 375, 2
\bibitem{}
De Lucia G., Helmi A., 2008, MNRAS, 391, 14
\bibitem{}
De Lucia G., Kauffmann G., White S. D. M., 2004, MNRAS, 349, 1101
\bibitem{}
De Lucia G., Tornatore L., Frenk C. S., Helmi A., Navarro J. F., White 
S. D. M., 2014, MNRAS, 445, 970
\bibitem{}
Desai V., Dalcanton J. J., Reed D., Mayer L., Quinn T., Governato F., 2004, 
MNRAS, 351, 265
\bibitem{}
Feltzing S., Eriksson K., Kleyna J., Wilkinson M. I., 2009, A\&A, 508, L1
\bibitem{}
Font A. S., et al., 2011, MNRAS, 417, 1260
\bibitem{}
Fulbright J. P., Rich R. M., Castro S., 2004, ApJ, 612, 447
\bibitem{}
Gibson B. K., 1994, MNRAS, 271, L35
\bibitem{}
Gilmore G., Norris J. E., Monaco L., Yong D., Wyse R. F. G., Geisler D., 2013, 
ApJ, 763, 61
\bibitem{}
Gratton R., Sneden C., Carretta E., 2004, ARA\&A, 42, 385
\bibitem{}
Hamuy M., 2003, ApJ, 582, 905
\bibitem{}
Hughes J., Wallerstein G., Bossi A., 2008, AJ, 136, 2321
\bibitem{}
Ishigaki M. N., Aoki W., Arimoto N., Okamoto S., preprint (arXiv:1401.1265)
\bibitem{}
Iwamoto K., Brachwitz F., Nomoto K., Kishimoto N., Umeda H., Hix W. R., 
Thielemann F.-K., 1999, ApJS, 125, 439
\bibitem{}
Karakas A. I., 2010, MNRAS, 403, 1413
\bibitem{}
Kauffmann G., Colberg J. M., Diaferio A., White S. D. M., 1999, MNRAS, 307, 529
\bibitem{}
Kennicutt R. C. Jr., 1998, ApJ, 498, 541 
\bibitem{}
Kirby E. N., et al., 2010, ApJS, 191, 352
\bibitem{}
Kirby E. N., Lanfranchi G. A., Simon J. D., Cohen J. G., Guhathakurta P., 2011, 
ApJ, 727, 78
\bibitem{}
Kobayashi C., Umeda H., Nomoto K., Tominaga N., Ohkubo T., 2006, ApJ, 653, 1145
\bibitem{}
Koch A., McWilliam A., Grebel E. K., Zucker D. B., Belokurov V., 2008, ApJ, 
688, L13
\bibitem{}
Komatsu E., et al., 2009, ApJS, 180, 330
\bibitem{}
Koposov S. E., et al., 2011, ApJ, 736, 146
\bibitem{}
Kroupa P., 2001, MNRAS, 322, 231
\bibitem{}
Lai D. K., Lee Y. S., Bolte M., Lucatello S., Beers T. C., Johnson J. A., 
Sivarani T., Rockosi C. M., 2011, ApJ, 738, 51
\bibitem{}
Lanfranchi G. A., Matteucci F., 2004, MNRAS, 351, 1338
\bibitem{}
Leaman R., 2012, AJ, 144, 183
\bibitem{}
Li Y.-S., De Lucia G., Helmi A., 2010, MNRAS, 401, 2036
\bibitem{}
Marcolini A., D'Ercole A., Brighenti F., Recchi S., 2006, MNRAS, 371, 643
\bibitem{}
Marconi G., Matteucci F., Tosi M., 1994, MNRAS, 270, 35
\bibitem{}
Martin C. L., Kobulnicky H. A., Heckman T. M., 2002, ApJ, 574, 663
\bibitem{}
Martin N. F., Ibata R. A., Chapman S. C., Irwin M., Lewis G. F., 2007, MNRAS, 
380, 281
\bibitem{}
Matteucci F., 2001, The chemical evolution of the Galaxy. Astrophysics and 
Space Science Library. Kluwer Academic Publishers, Dordrecht
\bibitem{}
Matteucci F., 2012, Chemical evolution of galaxies. Springer-Verlag Berlin, 
Heidelberg
\bibitem{}
Matteucci F., Greggio L., 1986, A\&A, 154, 279
\bibitem{}
Mayer L., Mastropietro C., Wadsley J., Stadel J., Moore B., 2006, MNRAS, 369, 
1021
\bibitem{}
McConnachie A. W., 2012, AJ, 144, 4
\bibitem{}
Melioli C., de Gouveia Dal Pino E. M., 2004, A\&A, 424, 817
\bibitem{}
Mori M., Burkert A., 2000, ApJ, 538, 559
\bibitem{}
Norris J. E., Gilmore G., Wyse R. F. G., Wilkinson M. I., Belokurov V., Evans 
N. W., Zucker D. B., 2008, ApJ, 689, L113
\bibitem{}
Norris J. E., Wyse R. F. G., Gilmore G., Yong D., Frebel A., Wilkinson M. I., 
Belokurov V., Zucker D. B., 2010a, ApJ, 723, 1632
\bibitem{}
Norris J. E., Yong D., Gilmore G., Wyse R. F. G., 2010b, ApJ, 711, 350
\bibitem{}
Oke J.~B., et al., 1995, PASP, 107, 375
\bibitem{}
Pagel B. E. J., 1997, Nucleosynthesis and Chemical Evolution of Galaxies.
Cambridge Univ. Press, Cambridge
\bibitem{}
Recchi S., 2014, Advances in Astronomy, Vol.~2014, Article ID~750754
\bibitem{}
Recchi S., Hensler G., 2013, A\&A, 551, A41
\bibitem{}
Recchi S., Matteucci F., D'Ercole A., 2001, MNRAS, 322, 800
\bibitem{}
Rieschick A., Hensler G., 2003, Ap\&SS, 284, 861
\bibitem{}
Romano D., Starkenburg E., 2013, MNRAS, 434, 471
\bibitem{}
Romano D., Karakas A. I., Tosi M., Matteucci F., 2010, A\&A, 522, A32
\bibitem{}
Roy J.-R., Kunth D., 1995, A\&A, 294, 432
\bibitem{}
Salpeter E. E., 1955, ApJ, 121, 161
\bibitem{}
Salvadori S., Ferrara A., 2009, MNRAS, 395, L6
\bibitem{}
Sawala T., et al., 2014, preprint (arXiv:1406.6362)
\bibitem{}
Sawala T., Scannapieco C., Maio U., White S., 2010, MNRAS, 402, 1599
\bibitem{}
Schmidt M., 1959, ApJ, 129, 243
\bibitem{}
Silich S. A., Tenorio-Tagle G., 1998, MNRAS, 299, 249
\bibitem{}
Springel V., White S. D. M., Tormen G., Kauffmann G., 2001, MNRAS, 328, 726
\bibitem{}
Springel V., Wang J., Vogelsberger M., Ludlow A., Jenkins A., Helmi A., Navarro 
J. F., Frenk C. S., White S. D. M., 2008, MNRAS, 391, 1685
\bibitem{}
Starkenburg E., et al., 2013, A\&A, 549, A88
\bibitem{}
Sutherland R. S., Dopita M. A., 1993, ApJS, 88, 253
\bibitem{}
Tenorio-Tagle G., 1996, AJ, 111, 1641
\bibitem{}
Tinsley B. M., 1980, Fundam. Cosm. Phys., 5, 287
\bibitem{}
Vader J. P., 1986, ApJ, 305, 669
\bibitem{}
Vader J. P., 1987, ApJ, 317, 128
\bibitem{}
van den Hoek L. B., Groenewegen M. A. T., 1997, A\&AS, 123, 305
\bibitem{}
Vargas L. C., Geha M., Kirby E. N., Simon J. D., 2013, ApJ, 767, 134
\bibitem{}
Vincenzo F., Matteucci F., Vattakunnel S., Lanfranchi G.~A., MNRAS, 441, 2815
\bibitem{}
Wang B., 1995, ApJ, 444, 590
\bibitem{}
Weidner C., Kroupa P., 2005, ApJ, 625, 754
\bibitem{}
Wolf J., Martinez G. D., Bullock J. S., Kaplinghat M., Geha M., Munoz R. R., 
Simon J. D., Avedo F. F., 2010, MNRAS, 406, 1220
\bibitem{}
Woosley S. E., Weaver T. A., 1995, ApJS, 101, 181
\bibitem{}
York D.~G., et al., 2000, AJ, 120, 1579
\end{thebibliography}
\end{document}